\documentclass[11pt,b4paper]{article}
\usepackage{bbm}
\usepackage{}
\usepackage{dsfont}
\usepackage{mathrsfs}
\usepackage{amssymb}
\usepackage{amsmath}
\usepackage{amsfonts}
\usepackage{amsthm}
\usepackage{epsfig}
\usepackage{enumerate}
\usepackage{graphicx}
\usepackage{epstopdf}
\usepackage{subfigure}
\usepackage{cite}

\newtheorem{lemma}{Lemma}

\newtheorem{assumption}{Assumption}

\long\def\comment#1{}

\title{Adaptive Fuzzy Tracking Control for Nonlinear State Constrained Pure-Feedback Systems With Input Delay via Dynamic Surface Technique
}
\author{Ju Wu\thanks{Research Institute of Intelligent Control and Systems, Harbin Institute of Technology, Harbin, 150080, China}, Tong Wang\thanks{Research Institute of Intelligent Control and Systems, Harbin Institute of Technology, Harbin, 150080, China}, Member, IEEE}

\date{}
\begin{document}

\maketitle \thispagestyle{plain}


\begin{abstract}
This brief constructs the adaptive backstepping control scheme for a class of pure-feedback systems with input delay and full state constraints. With the help of Mean Value Theorem, the pure-feedback system is transformed into strict-feedback one. Barrier Lyapunov functions are employed to guarantee all of the states remain constrained within predefined sets. By introducing the Pade approximation method and corresponding intermediate, the impact generated by input delay on the output tracking performance of the system can be eliminated. Furthermore, a low-pass filter driven by a newly-defined control input, is employed to generate the actual control input, which facilitates the design of backstepping control. To approximate the unknown functions with a desired level of accuracy, the fuzzy logic systems (FLSs) are utilized by choosing appropriate fuzzy rules, logics and so on. The minimal learning parameter (MLP) technique is employed to decrease the number of nodes and parameters in FLSs, and dynamic surface control (DSC) technique is leveraged to avoid so-called "explosion of complexity". Moreover, smooth robust compensators are introduced to circumvent the influences of external disturbance and approximation errors. By stability analysis, it is proved that all of signals in the closed-loop system are semi-globally ultimately uniform bounded, and the tracking error can be within a arbitrary small neighbor of origin via selecting appropriate parameters of controllers. Finally, the results of numerical illustration are provided to demonstrate the effectiveness of the designed method.


\end{abstract}



\section{Introduction}
Adaptive control emerged in the 1950s to address the limitations of traditional control systems, such as constant-gain feedback, which were insufficient for handling uncertainties and dynamic variations in systems like supersonic aircraft and industrial processes. Its importance lies in its ability to modify controller behavior in response to changes in system dynamics and disturbances, with seminal frameworks like Model Reference Adaptive Control (MRAC) and Self-Tuning Regulators (STR) becoming foundational tools for achieving stability, robustness, and adaptability across diverse applications \cite{aastrom2021history}. The shift from linear to nonlinear adaptive control addresses the limitations of linear growth constraints and unrealistic full-state feedback assumptions, enabling the control of complex nonlinear systems critical for practical applications like robotics, electric motors, and automotive suspensions \cite{kanellakopoulos1992adaptive}. The emergence of backstepping methods revolutionized adaptive control for nonlinear systems by introducing a recursive design framework capable of systematically handling severe nonlinearities and uncertainties \cite{10.5555/546475}.

Fuzzy logic systems and neural networks have been widely used for approximation of unknown nonlinear dynamics in adaptive control. \cite{258721} first proved fuzzy systems to be universal approximators, using Stone-Weierstrass theorem. The foundational works by \cite{cybenko1989approximation} and \cite{HORNIK1989359} proved the inherent potential of neural networks as universal approximators, demonstrating their capability to approximate any continuous function on a compact set to arbitrary precision. \cite{227383} developed a direct adaptive fuzzy controller that ensures global stability and uniform boundedness without requiring an accurate system model, and incorporated fuzzy control rules for faster adaptation. The unified framework of adaptive backstepping control design, comprised of both adaptive fuzzy state feedback and observer-based output feedback control design schemes was proposed for the single-input and single-output (SISO) uncertain non-strict feedback nonlinear systems \cite{7429731}. Layered neural networks (NNs) were employed to identify unknown nonlinear functions in a feedback-linearizable discrete-time system \cite{chen1992adaptive}. An adaptive backstepping control scheme for strict-feedback nonlinear systems using NNs for approximation was developed with a modified Lyapunov function, supporting explicit transient performance tuning and parallel processing through structural properties \cite{zhang2000adaptive}. \cite{wang2002adaptive} applied adaptive backstepping with NNs-based design to a broader class of SISO uncertain nonlinear systems. To overcome the difficulty introduced by triangular structure of nonlinear pure-feedback systems with unknown nonlinear functions, implicit function theorem was firstly exploited to assert the existence of the continuous desired virtual controls \cite{ge2002adaptive}. The predefined performance widely found in engineering application requirements has been discussed for adaptive control of nonlinear systems. For strict-feedback systems with unknown nonlinearities, \cite{zhang2017fuzzy} developed a low-complexity adaptive fuzzy output feedback control scheme to achieve finite-time output error convergence, \cite{wang2022adaptive} achieved predefined time and accuracy stability, ensuring practical applicability with adjustable convergence time and accuracy. 

To address the "explosion of complexity" of the adaptive backstepping algorithm for nonlinear uncertain systems, first-order low-pass filters were introduced to avoid differentiation of model nonlinearities \cite{yip1998adaptive}. An improved adaptive Dynamic Surface Control (DSC) approach was proposed, using nonlinear adaptive filters instead of first-order low-pass filters, with stability ensured by novel Lyapunov functions incorporating flat zones \cite{zhang2018improved}. DSC and backstepping-based adaptive control design were explored for strict-feedback nonlinear systems in \cite{wang2005neural} and for pure-feedback nonlinear systems in \cite{liu2015adaptive}. The issue of unknown control direction for strict-feedback system was addressed using the Nussbaum function in \cite{ma2018adaptive}, while \cite{wang2019adaptive} addressed the adaptive control of nonstrict-feedback nonlinear systems with unknown virtual control coefficients by integrating a DSC scheme with the Nussbaum gain technique.

Constraints are inherent in nearly all physical systems, making the effective management of constraints in control design a critical topic to prevent performance degradation, both in practice and theory. Invariant sets provided a foundational framework for addressing state and input constraints in linear systems, with applications extending to robustness analysis and state-feedback control \cite{blanchini1999set}. \cite{dehaan2005extremum} developed an adaptive extremum-seeking control method, leveraging interior-point barrier functions, to ensure feasibility of state constraints and achieve convergence to the minimizer of an objective function with unknown parameters. \cite{krstic2006nonovershooting} introduced a method for achieving nonovershooting output tracking in SISO strict-feedback nonlinear systems, ensuring trajectories are tracked "from below" for arbitrary initial conditions. \cite{tee2009control} first developed a control strategy for state-constrained nonlinear systems in strict-feedback form using a Barrier Lyapunov Function (BLF), which ensures state constraints are not violated by growing unbounded as states approach their limits. BLFs-based adaptive control was employed to address a class of constrained nonlinear systems, including SISO and pure-feedback systems \cite{liu2016barrier}, with practical requirements such as full-state, output \cite{tee2009barrier}, state-and-time-dependent constraints \cite{liu2021adaptive}, and unknown control directions \cite{liu2017barrier}. To simplify feasibility checks, \cite{tee2012control} introduced Integral Barrier Lyapunov Functions (iBLF) to reduce conservatism by integrating state constraints with error dynamics. Variants of iBLFs-based adaptive NN control have been developed to address a class of uncertain and constrained nonlinear systems, including MIMO systems in block-triangular form \cite{liu2016adaptive} and stochastic systems with symmetric and asymmetric full-state constraints \cite{gao2021iblf}. Under actuator saturation, a modified reference model was employed to enable correct adaptation for NN-based adaptive control \cite{946179}, and \cite{5723705} developed an adaptive backstepping control with a Nussbaum function to address input saturation for a class of single input uncertain nonlinear systems.

Note that the control problem of time delay occurs in most of practical systems and often leads to performance degradation, thus driven by the need of eliminating ill-effects of time delay in systems, this control problem has attracted remarkable attention. \cite{jankovic2001control} as a benchmark work for solving time-delay problem introduced control Lyapunov-Razumikhin to construct robust stabilizing control laws for time delay systems. \cite{wang2008adaptive} investigated adaptive fuzzy controller for a class of uncertain SISO nonlinear time-delay systems in strict-feedback form, which was extended to nonaffine form by \cite{wang2010approximation}. \cite{wu2017adaptive}, \cite{li2016adaptive}, \cite{li2018neural} used Pade approximation to tackle with input delay mainly induced by network in the data transmitting process. 

Up to now, there are few work handling the tracking control problem for the state constrained pure-feedback nonlinear systems with input delay. Thus, in this paper, we try to construct a novel adaptive fuzzy tracking control approach for a class of pure-feedback nonlinear systems with state constraints and input delay.

The rest of this paper is organized as follows.
Section II gives the problem formulation and preliminaries.
The novel adaptive fuzzy tracking controller design is
given in Section III. Section IV presents Feasibility check. The simulation example is presented in
Section V to show the effectiveness of the proposed control
scheme. Finally, Section VI concludes this paper.

\section{Problem Statement and Preliminaries}
Considering the following pure-feedback system with input delay and full state constraints
\begin{equation}
{ \left\{ \begin{array}{l}
{{{\dot{x}}}_{i}}={{f}_{i}}({{{\bar{x}}}_{i}},{{x}_{i+1}})+{{d}_{i}}(t),\text{       }i=1,2,\ldots ,n-1, \\
{{{\dot{x}}}_{n}}={{f}_{n}}({{{\bar{x}}}_{n}},u(t-\tau ))+{{d}_{n}}(t),\\
y={{x}_{1}}
\end{array} \right.} \label{pf plant}
\end{equation}
where $\bar x_i=[x_1,x_2,\cdots,x_i]^T \in \mathbb{R}^i, i=1,2,\cdots,n, (x=\bar x_n),$ are the state vectors. $u(t) \in \mathbb{R}, y(t) \in \mathbb{R},$ and ${{d}_{i}}(t) \in \mathbb{R}$ are the input, measured output, and bounded unknown external disturbances of (\ref{pf plant}), respectively. The state should remain within the constraints $\vert x_i \vert \le {k_{ci}}, i=1,2, \ldots ,n$, where ${k_{ci}}$ are positive constants. In addition, the following inequality holds for the external disturbances $\vert {{d}_{i}}(t) \vert \le {{d}_{iM}}, {{f}_{i}}({{{\bar{x}}}_{i}},{{x}_{i+1}})$ are the unknown smooth nonlinear functions, $\tau$ denotes the constant input delay. $y_d$ is the reference output signal in this paper.

The object of this paper is to design an appropriate controller to guarantee that all of the signals in the closed-loop system are bounded, the states remain within the constraints, and the output tracking error converges to an arbitrarily small neighbor of origin.

To tackle the problem of input delay, Pade approximation approach is introduced. we have
\begin{align}
L\left\{ u(t-\tau ) \right\}=\exp (-\tau s)L\left\{ u(t) \right\}=\frac{\exp (-\tau s/2)}{\exp (\tau s/2)}L\left\{ u(t) \right\} \nonumber\\
\approx \frac{(1-\tau s/2)}{(1+\tau s/2)}L\left\{ u(t) \right\}\,\label{pade}
\end{align}
where $L\left\{ u(t) \right\}$ is the Laplace transform of $u(t)$, and $s$ is the Laplace variable. For further investigation, the intermediate variable $\chi$ is introduced. Then, we have
\begin{align}
\frac{1-\tau s/2}{1+\tau s/2}L\left\{ u(t) \right\}=L\left\{ \chi (t) \right\}-L\left\{ u(t) \right\},\label{Xn_1}
\end{align}
by inverse Laplace transform, (\ref{Xn_1}) can be written as
\begin{align}
\dot{\chi }=-\lambda \chi +2\lambda u(t).\label{Xn_2}
\end{align}
where $\lambda=2/{\tau}$. To circumvent the difficulties caused by non-affine systems in controller design, let the control input $u(t)$ generated by a low-pass filter driven by a control input $v(t)$
\begin{align}
\dot{u}(t)=-\kappa u(t)+v(t),\label{lpf}
\end{align}
where $\kappa$ is the designed constant for the low-pass filter. Therefore, the original system (\ref{pf plant}) can be rewritten as
\begin{equation}
{ \left\{ \begin{array}{l}
{{{\dot{x}}}_{i}}={{f}_{i}}({{{\bar{x}}}_{i}},{{x}_{i+1}})+{{d}_{i}}(t),\text{       }i=1,2,\ldots ,n-1, \\
{{\dot{x}}_{n}}={{f}_{n}}({{\bar{x}}_{n}},\chi (t)-u(t))+{{d}_{n}}(t) \\
\dot{\chi }=-\lambda\chi +2\lambda u(t) \\
\dot{u}(t)=-\kappa u(t)+v(t) \\
y={{x}_{1}}
\end{array} \right.} \label{Xn 3}
\end{equation}
Next, the following commonly found assumption are introduced

\begin{assumption}
For the control of pure feedback system (\ref{pf plant}), define ${{g}_{i}}({{\bar{x}}_{i}},{{x}_{i+1}})=\frac{\partial {{f}_{i}}({{{\bar{x}}}_{i}},{{x}_{i+1}})}{\partial {{x}_{i+1}}}, i=1,2, \ldots ,n$. The sign of ${{g}_{i}}({{\bar{x}}_{i}},{{x}_{i+1}})$ are known, and there exist unknown constants $g_{i0}$ and $gi1$ such that ${{g}_{i0}} \le \vert{{g}_{i}}(\cdot )\vert\le {{g}_{i1}}, \forall x\in {{\Omega }_{x}}\subset \mathbb{R}^n$. Without loss of generality, we shall assume that ${{g}_{i0}} \le {{g}_{i}}(\cdot ), \forall x\in {{\Omega }_{x}}\subset \mathbb{R}^n$
\end{assumption}
\begin{assumption}
There exist the known constants $A_{0},C_{0}$, such that the desired trajectory $y_{d}$ satisfies $\vert y_{d} \vert \le A_{0} \le k_{c1}$ and there is a compact set ${{\Omega }_{0}}=\{{{[{{y}_{d}},{{\dot{y}}_{d}},{{\ddot{y}}_{d}}]}^{T}}:y_{d}^{2}+\dot{y}_{d}^{2}+\ddot{y}_{d}^{2}\le {{C}_{0}}\}$.
\end{assumption}
\begin{assumption}
For $1 \le i \le n$, there exist unknown positive constant $d_{iM}$, satisfying $\vert {d_{i}}(t) \vert \le d_{iM}$
\end{assumption}

\begin{lemma}
\cite{ge2002adaptive} Assume that $f(x,u):\mathbb{R}^n \times \mathbb{R} \to \mathbb{R}$ is continuously differentiable $\forall (x,u)\in \mathbb{R}^n \times \mathbb{R}$, and there exists a positive constant $d$, such that $\partial f(x,u)/\partial u > d >0, \forall (x,u)\in \mathbb{R}^n \times \mathbb{R}$. Then there exists a continue(smooth) function ${{u}^{*}}=u(x)$ satisfying that $f(x,{u}^{*})=0$.
\end{lemma}

\begin{lemma}
Note that the hyperbolic tangent function $\tanh (\cdot )$ is continuous and differentiable, it fulfills that for any $p \in \mathbb{R}$ and $\upsilon > 0$, the following inequalities can be satisfied
\begin{align}
0\le \left| p \right|-p\tanh \left( \frac{p}{\upsilon } \right)<0.2785\upsilon
\end{align}
\begin{align}
0\le p\tanh \left( \frac{p}{\upsilon } \right).
\end{align}
\end{lemma}

\begin{lemma}
For a continuous function $\psi (x):\mathbb{R}^n \to \mathbb{R}$ which is defined on a compact ${{\Omega }_{x}}\in \mathbb{R}^n$, there exists a fuzzy logic system ${W}^{T}\xi (x)$ which can be used to approximate $\psi (x)$ with the technique including singleton, center average defuzzification and product inference, satisfying that
\begin{align}
\psi (x)={W}^{T}\xi (x)+&\varepsilon \\
\underset{x\in {{\Omega }_{x}}}{\mathop{\sup }}\,\left| \psi (x)-{W}^{T}\xi (x) \right|&\le {{\varepsilon }^{*}}
\end{align}
where $W={{[{{\omega }_{1}},{{\omega }_{2}},\ldots ,{{\omega }_{N}}]}^{T}}$ is the ideal weight vector. $\xi(x)$ and $\zeta(x)$ are basic functions and Gaussian functions respectively, which can be expressed as
\begin{align}
\xi (x)&=\frac{{{[{{\zeta }_{1}}(x),{{\zeta }_{2}}(x),...,{{\zeta }_{N}}(x)]}^{T}}}{\sum\limits_{j=1}^{N}{{{\zeta }_{j}}(x)}}, \\
{{\zeta }_{j}}(x)&=\exp \left( \frac{-{{(x-{{l}_{j}})}^{T}}(x-{{l}_{j}})}{\eta _{j}^{T}{{\eta }_{j}}} \right)
\end{align}
where ${{l}_{j}}={{[{{l}_{j1}},{{l}_{j2}},\ldots ,{{l}_{jn}}]}^{T}}$ is the center vector, ${{\eta }_{j}}={{[{{\eta }_{j1}},{{\eta }_{j2}},\ldots ,{{\eta }_{jn}}]}^{T}}$ is the width of Gaussian function, while $n$ and $N$ are the number of system input and rules of fuzzy logic systems respectively.
\end{lemma}

\section{Controller Design}
In this section, backstepping technique is employed to facilitate adaptive fuzzy controller design for system $(\ref{pf plant})$. We have the following changes of coordinates
\begin{equation}
{ \left\{ \begin{array}{l}
{{z}_{1}}={{x}_{1}}-{{w}_{1}},\nonumber \\
{{z}_{i}}={{x}_{i}}-{{w}_{i}}, i=1,2,\ldots ,n, \\
{{z}_{n+1}}=\chi (t)-u(t)-{{w}_{n+1}}, \nonumber
\end{array} \right.} \label{pf plant3}
\end{equation}
where ${z_{1}}$ is the tracking error, $w_1=y_d$, and ${{w}_{i}}$ is the output of the first-order filter with intermediate ${\alpha_{i}}$ as the input on the basis of DSC technique. At the first $n$ steps, the virtual control ${\alpha_{i}}$ is designed to guarantee the corresponding subsystem toward equilibrium condition. To tackle the problem of input time delay, control input $v(t)$ generating actual system input $u(t)$ through a low-pass filter is designed at step $n+1$.

\textbf{Step $1$ :} Considering the time derivative of ${z_{1}}$
\begin{align}
{{\dot{z}}_{1}}={{f}_{1}}({{x}_{1}},{{x}_{2}})+{{d}_{1}}(t)-{{\dot{w}}_{1}}.\label{z1}
\end{align}
From the Assumption 1, we know that $\partial {f_{1}}(x_1,x_2)/\partial x_2 > g_10 >0, \forall (x_1,x_2)\in \mathbb{R}^2$. Define $v_1=-\dot{w}_1$, we have the following inequality
\begin{align}
\frac{\partial {{f}_{1}}({{x}_{1}},{{x}_{2}})+{{v}_{1}}}{\partial {{x}_{2}}}>{{g}_{10}}>0
\end{align}
According ro Lemma 1, there exists smooth ideal input $x_2=\alpha _{1}^{*}(x_1,v_1), \forall (x_1,v_1)\in \mathbb{R}^2$, such that
\begin{align}
{{f}_{1}}({{x}_{1}},\alpha _{1}^{*})+{{v}_{1}}=0.
\end{align}
By Mean Value Theorem, there exists ${\lambda_{1}} (0<\lambda_1<1)$ satisfying
\begin{align}
{{f}_{1}}({{x}_{1}},{{x}_{2}})={{f}_{1}}({{x}_{1}},\alpha _{1}^{*})+{{g}_{1{{\lambda }_{1}}}}({{x}_{2}}-\alpha _{1}^{*}),\label{mth1}
\end{align}
where ${{g}_{1{{\lambda }_{1}}}}={{g}_{1}}({{x}_{1}},{{x}_{2{{\lambda }_{1}}}}), {{x}_{2{{\lambda }_{1}}}}={{\lambda }_{1}}{{x}_{2}}+(1-{{\lambda }_{1}})\alpha _{1}^{*}$. Substitute $(\ref{mth1})$ into $(\ref{z1})$, we yield
\begin{align}
{{\dot{z}}_{1}}={{g}_{1{{\lambda }_{1}}}}({{x}_{2}}-\alpha _{1}^{*})+{{d}_{1}}(t). \label{dz1}
\end{align}

Define the Barrier Lyapunov function as follows
\begin{align}
{{V}_{z1}}=\frac{1}{2{{g}_{1{{\lambda }_{1}}}}}\ln \left( \frac{k_{b1}^{2}}{k_{b1}^{2}-z_{1}^{2}} \right),\label{BLF1}
\end{align}
where $k_{b1} \le k_{c1}-{A}_{0}, {A}_{0}$ is the bound of the desired trajectory $y_d$, then the time derivative of (\ref{BLF1}) is
\begin{align}
{{\dot{V}}_{z1}}=\frac{1}{{{g}_{1{{\lambda }_{1}}}}}\left( \frac{{{z}_{1}}{{{\dot{z}}}_{1}}}{k_{b1}^{2}-z_{1}^{2}} \right)-\frac{{{{\dot{g}}}_{1{{\lambda }_{1}}}}}{2{{g}_{1{{\lambda }_{1}}}}^{2}}\ln \left( \frac{k_{b1}^{2}}{k_{b1}^{2}-z_{1}^{2}} \right).\label{dBLF1}
\end{align}
By substituting (\ref{dz1}) into (\ref{dBLF1}), we obtain
\begin{align}
{{\dot{V}}_{z1}}=\frac{{{z}_{1}}({{x}_{2}}-\alpha _{1}^{*})}{k_{b1}^{2}-z_{1}^{2}}+\frac{1}{{{g}_{1{{\lambda }_{1}}}}}\frac{{{z}_{1}}{{d}_{1}}(t)}{k_{b1}^{2}-z_{1}^{2}}-\frac{{{{\dot{g}}}_{1{{\lambda }_{1}}}}}{2{{g}_{1{{\lambda }_{1}}}}^{2}}\ln \left( \frac{k_{b1}^{2}}{k_{b1}^{2}-z_{1}^{2}} \right). \label{dBLF2}
\end{align}
According to Lemma 3, $\alpha _{1}^{*}$ can be approximated by appropriate FLSs as follows
\begin{align}
\alpha _{1}^{*}({{Z}_{1}})={{W}_{1}}^{T}{{\xi }_{1}}({{Z}_{1}})+{{\varepsilon }_{1}}({{Z}_{1}}), \label{FA1}
\end{align}
where ${{Z}_{1}}={{\left[ {{x}_{1}},{{{\dot{w}}}_{1}} \right]}^{T}}\in {{R}^{2}}$ is the argument vector of unknown function $\alpha _{1}^{*}({{x}_{1}},{{{\dot{w}}}_{1}})$. And there exists unknown constant ${{\varepsilon }_{1}}^{*}$, such that $\vert {{\varepsilon }_{1}} \vert \le {{\varepsilon }_{1}}^{*}$. Then, substituting (\ref{FA1}) into (\ref{dBLF2}) yields
\begin{align}
{{\dot{V}}_{z1}}=\frac{{{z}_{1}}({{z}_{2}}+{{e}_{2}}+{{\alpha }_{1}})}{k_{b1}^{2}-z_{1}^{2}}+\frac{1}{{{g}_{1{{\lambda }_{1}}}}}\frac{{{z}_{1}}{{d}_{1}}(t)}{k_{b1}^{2}-z_{1}^{2}}-\frac{{{z}_{1}}\alpha _{1}^{*}}{k_{b1}^{2}-z_{1}^{2}}-\frac{{{{\dot{g}}}_{1{{\lambda }_{1}}}}}{2{{g}_{1{{\lambda }_{1}}}}^{2}}\ln \left( \frac{k_{b1}^{2}}{k_{b1}^{2}-z_{1}^{2}} \right),\label{dBLF3}
\end{align}
where $e_{2}=w_{2}-\alpha_{1}, x_{2}=z_{2}+e_{2}+\alpha_{1}$. By using Young's inequality, we have
\begin{align}
-\frac{{{z}_{1}}\alpha _{1}^{*}}{k_{b1}^{2}-z_{1}^{2}}&\le \left| \frac{{{z}_{1}}}{k_{b1}^{2}-z_{1}^{2}} \right|\left( |{{W}_{1}}^{T}\xi {}_{1}|+|{{\varepsilon }_{1}}| \right) \nonumber \\
&\le \frac{{{z}_{1}}^{2}{{\left\| {{W}_{1}} \right\|}^{2}}\xi _{1}^{T}\xi {}_{1}}{2{{a}_{\text{1}}}^{2}{{(k_{b1}^{2}-z_{1}^{2})}^{2}}}+\frac{{{a}_{\text{1}}}^{2}}{2}+\left| \frac{{{z}_{1}}}{k_{b1}^{2}-z_{1}^{2}} \right|{{\varepsilon }_{1}}^{*},\label{YI1}
\end{align}
where $a_1>0$ is a designed constant. By substituting (\ref{YI1}) into (\ref{dBLF3}), we have
\begin{align}
{{{\dot{V}}}_{z1}}\le& \frac{{{z}_{1}}({{z}_{2}}+{{e}_{2}})}{k_{b1}^{2}-z_{1}^{2}}+\frac{{{z}_{1}}{{\alpha }_{1}}}{k_{b1}^{2}-z_{1}^{2}}+\frac{1}{{{g}_{10}}}\left| \frac{{{z}_{1}}}{k_{b1}^{2}-z_{1}^{2}} \right|\left| {{d}_{1}}(t) \right|+\frac{{{z}_{1}}^{2}{{\left\| {{W}_{1}} \right\|}^{2}}\xi _{1}^{T}{{\xi }_{1}}}{2{{a}_{\text{1}}}^{2}{{(k_{b1}^{2}-z_{1}^{2})}^{2}}} \nonumber \\
&+\frac{{{a}_{\text{1}}}^{2}}{2}+\left| \frac{{{z}_{1}}}{k_{b1}^{2}-z_{1}^{2}} \right|{{\varepsilon }_{1}}^{*}-\frac{{{{\dot{g}}}_{1{{\lambda }_{1}}}}}{2{{g}_{1{{\lambda }_{1}}}}^{2}}\ln (\frac{k_{b1}^{2}}{k_{b1}^{2}-z_{1}^{2}}) \nonumber  \\
\le& \frac{{{z}_{1}}({{z}_{2}}+{{e}_{2}})}{k_{b1}^{2}-z_{1}^{2}}+\frac{{{z}_{1}}{{\alpha }_{1}}}{k_{b1}^{2}-z_{1}^{2}}+\frac{{{z}_{1}}^{2}{{\theta }_{1}}\xi _{1}^{T}{{\xi }_{1}}}{{{(k_{b1}^{2}-z_{1}^{2})}^{2}}}+\left| \frac{{{z}_{1}}}{k_{b1}^{2}-z_{1}^{2}} \right|{{\delta }_{1}}^{*} \nonumber \\
&+\frac{{{a}_{\text{1}}}^{2}}{2}-\frac{{{{\dot{g}}}_{1{{\lambda}_{1}}}}}{2{{g}_{1{{\lambda
}_{1}}}}^{2}}\ln(\frac{k_{b1}^{2}}{k_{b1}^{2}-z_{1}^{2}}),\label{dBLF4}
\end{align}
where ${{\delta }_{1}}^{*}=\frac{{{d}_{1M}}}{{{g}_{10}}}+{{\varepsilon }_{1}}^{*},\text{ }{{\theta }_{1}}=\frac{{{\left\| {{W}_{1}} \right\|}^{2}}}{2{{a}_{1}}^{2}}$. By using Young's inequality we have
\begin{align}
\frac{{{z}_{1}}{{e}_{2}}}{k_{b1}^{2}-z_{1}^{2}}\le \frac{{{z}_{1}}^{2}}{{{\left( k_{b1}^{2}-z_{1}^{2} \right)}^{2}}}+\frac{{{e}_{2}}^{2}}{4}. \label{YI2}
\end{align}
By substituting (\ref{YI2}) into (\ref{dBLF4}), we obtain
\begin{align}
{{\dot{V}}_{z1}}\le&\frac{{{z}_{1}}{{z}_{2}}}{k_{b1}^{2}-z_{1}^{2}}+\frac{{{z}_{1}}{{\alpha }_{1}}}{k_{b1}^{2}-z_{1}^{2}}+\frac{{{z}_{1}}^{2}{{\theta }_{1}}\xi _{1}^{T}{{\xi }_{1}}}{{{(k_{b1}^{2}-z_{1}^{2})}^{2}}}+\frac{{{z}_{1}}^{2}}{{{(k_{b1}^{2}-z_{1}^{2})}^{2}}} \nonumber \\
&+\frac{e_{2}^{2}}{4}+\left| \frac{{{z}_{1}}}{k_{b1}^{2}-z_{1}^{2}} \right|{{\delta }_{1}}^{*}+\frac{a_{1}^{2}}{2}+{{\mu }_{1}}({{\Gamma }_{1}}),\label{dBLF5}
\end{align}
where ${{\mu }_{1}}({{\Gamma }_{1}})$ is a positive continuous function, and ${{\Gamma }_{1}}=[{{z}_{1}},{{y}_{d}},{{\dot{y}}_{d}}]^{T}\in \mathbb{R}^{3}$, satisfying $\left| \frac{{{{\dot{g}}}_{1{{\lambda }_{1}}}}}{2{{g}_{1{{\lambda }_{1}}}}^{2}}\ln \left( \frac{k_{b1}^{2}}{k_{b1}^{2}-z_{1}^{2}} \right) \right|\le {{\mu }_{1}}({{\Gamma }_{1}})$.

Then, design the virtual control law $\alpha_{1}$ and adaptation laws as follows
\begin{align}
{{\alpha }_{1}}&=-{{K}_{1}}{{z}_{1}}-\frac{{{z}_{1}}{{{\hat{\theta }}}_{1}}\xi _{1}^{T}{{\xi }_{1}}}{k_{b1}^{2}-z_{1}^{2}}-{{\hat{\delta }}_{1}}\tanh \left( \frac{\left( \frac{{{z}_{1}}}{k_{b1}^{2}-z_{1}^{2}} \right)}{{{\upsilon }_{1}}} \right)-\frac{{{z}_{1}}}{k_{b1}^{2}-z_{1}^{2}},\label{VI1} \\
{{\dot{\hat{\delta }}}_{1}}&={{\gamma }_{1}}\frac{{{z}_{1}}}{k_{b1}^{2}-z_{1}^{2}}\tanh \left( \frac{\left( \frac{{{z}_{1}}}{k_{b1}^{2}-z_{1}^{2}} \right)}{{{\upsilon }_{1}}} \right)-{{\sigma }_{1}}{{\gamma }_{1}}{{\hat{\delta }}_{1}},\label{AL1} \\
{{\dot{\hat{\theta }}}_{1}}&={{\beta }_{1}}\frac{{{z}_{1}}^{2}\xi _{1}^{T}{{\xi }_{1}}}{k_{b1}^{2}-z_{1}^{2}}-{{\sigma }_{1}}{{\beta }_{1}}{{\hat{\theta }}_{1}},\label{dAL1}
\end{align}
where ${{K}_{1}}>0,{{\beta }_{1}}>0,{{\gamma }_{1}}>0,{{\sigma }_{1}}>0,{{\upsilon }_{1}}>0$ are the designed parameters, and the term ${{\hat{\theta }}_{1}},{{\hat{\delta }}_{1}}$ are the estimates of ${{\theta }_{1}}$ and ${{\delta }_{1}^{*}}$ respectively. The term ${{\hat{\delta }}_{1}}$ is used as a robust compensator to deal with the problem of external disturbance and approximation error. Note that if we choose ${{\hat{\theta }}_{1}}(0)>0, {{\hat{\delta }}_{1}}(0)>0$, then ${{\hat{\theta }}_{1}}(t)>0, {{\hat{\delta }}_{1}}(t)>0, \forall t>0$.

Considering the following augmented Lyapunov function
\begin{align}
{{V}_{1}}=\frac{1}{2{{g}_{1{{\lambda }_{1}}}}}\ln \left( \frac{k_{b1}^{2}}{k_{b1}^{2}-z_{1}^{2}} \right)\text{+}\frac{1}{2{{\gamma }_{1}}}{{\tilde{\delta }}_{1}}^{2}+\frac{1}{2{{\beta }_{1}}}{{\tilde{\theta }}_{1}}^{2},\label{NLF1}
\end{align}
where ${{\tilde{\delta }}_{1}}\text{=}\delta _{1}^{*}\text{-}{{\hat{\delta }}_{1}},{{\tilde{\theta }}_{1}}\text{=}{{\theta }_{1}}\text{-}{{\hat{\theta }}_{1}}$. Then the time derivative of (\ref{NLF1}) is
\begin{align}
{{{\dot{V}}}_{1}}\le&\frac{{{z}_{1}}{{z}_{2}}}{k_{b1}^{2}-z_{1}^{2}}+\frac{{{z}_{1}}{{\alpha }_{1}}}{k_{b1}^{2}-z_{1}^{2}}+\frac{{{z}_{1}}^{2}{{\theta }_{1}}\xi _{1}^{T}{{\xi }_{1}}}{{{(k_{b1}^{2}-z_{1}^{2})}^{2}}}+\frac{{{z}_{1}}^{2}}{{{(k_{b1}^{2}-z_{1}^{2})}^{2}}}+\frac{{{e}_{2}}^{2}}{4} \nonumber \\
&+\left| \frac{{{z}_{1}}}{k_{b1}^{2}-z_{1}^{2}} \right|{{\delta }_{1}}^{*}+\frac{a_{1}^{2}}{2}+{{\mu }_{1}}({{\Gamma }_{1}})-\frac{1}{{{\gamma }_{1}}}{{{\tilde{\delta }}}_{1}}{{{\dot{\hat{\delta }}}}_{1}}-\frac{1}{{{\beta }_{1}}}{{{\tilde{\theta }}}_{1}}{{{\dot{\hat{\theta }}}}_{1}}.\label{dNLF1}
\end{align}
By substituting (\ref{VI1}),(\ref{AL1}),(\ref{dAL1}) into (\ref{dNLF1}), we obtain
\begin{align}
{{{\dot{V}}}_{1}}\le&\frac{{{z}_{1}}{{z}_{2}}}{k_{b1}^{2}-z_{1}^{2}}+\frac{{{z}_{1}}{{\alpha}_{1}}}{k_{b1}^{2}-z_{1}^{2}}+\frac{{{z}_{1}}^{2}{{\theta }_{1}}\xi _{1}^{T}{{\xi }_{1}}}{{{(k_{b1}^{2}-z_{1}^{2})}^{2}}}+\frac{e_{2}^{2}}{4}+\left| \frac{{{z}_{1}}}{k_{b1}^{2}-z_{1}^{2}} \right|{{\delta }_{1}}^{*}+\frac{{{a}_{1}}^{2}}{2}+{{\mu }_{1}}({{\Gamma }_{1}})-\frac{1}{{{\gamma }_{1}}}{{\tilde{\delta }}_{1}}{{\dot{\hat{\delta }}}_{1}}-\frac{1}{{{\beta }_{1}}}{{\tilde{\theta }}_{1}}{{\dot{\hat{\theta }}}_{1}} \nonumber \\
\le&\frac{{{z}_{1}}{{z}_{2}}}{k_{b1}^{2}-z_{1}^{2}}-\frac{{{K}_{1}}{{z}_{1}}^{2}}{k_{b1}^{2}-z_{1}^{2}}+\frac{e_{2}^{2}}{4}+\left| \frac{{{z}_{1}}}{k_{b1}^{2}-z_{1}^{2}} \right|{{\delta }_{1}}^{*}-\frac{{{z}_{1}}}{k_{b1}^{2}-z_{1}^{2}}{{{\hat{\delta }}}_{1}}\tanh \left( \frac{\left( \frac{{{z}_{1}}}{k_{b1}^{2}-z_{1}^{2}} \right)}{{{\upsilon }_{1}}} \right) \nonumber \\
&+\frac{{{a}_{1}}^{2}}{2}+{{\mu }_{1}}({{\Gamma }_{1}})+{{{\tilde{\delta }}}_{1}}\left( {{\sigma }_{1}}{{{\hat{\delta }}}_{1}}-\frac{{{z}_{1}}}{k_{b1}^{2}-z_{1}^{2}}\tanh \left( \frac{\left( \frac{{{z}_{1}}}{k_{b1}^{2}-z_{1}^{2}} \right)}{{{\upsilon }_{1}}} \right) \right)+{{\sigma }_{1}}{{{\tilde{\theta }}}_{1}}{{{\hat{\theta }}}_{1}} \nonumber \\
\le&-\frac{{{K}_{1}}{{z}_{1}}^{2}}{k_{b1}^{2}-z_{1}^{2}}\text{+}\frac{{{z}_{1}}{{z}_{2}}}{k_{b1}^{2}-z_{1}^{2}}+\frac{{{e}_{2}}^{2}}{4}+\frac{{{a}_{1}}^{2}}{2}+{{\mu }_{1}}({{\Gamma }_{1}}) \nonumber \\
&+{{\delta }_{1}}^{*}\left( \left| \frac{{{z}_{1}}}{k_{b1}^{2}-z_{1}^{2}} \right|-\frac{{{z}_{1}}}{k_{b1}^{2}-z_{1}^{2}}\tanh \left( \frac{\left( \frac{{{z}_{1}}}{k_{b1}^{2}-z_{1}^{2}} \right)}{{{\upsilon }_{1}}} \right) \right) \nonumber \\
&+{{\sigma }_{1}}({{{\tilde{\theta }}}_{1}}{{{\hat{\theta }}}_{1}}+{{{\tilde{\delta }}}_{1}}{{{\hat{\delta }}}_{1}}).\label{dNLF2}
\end{align}
According to Lemma 2, we obtain
\begin{align}
{{\delta }_{1}}^{*}\left( \left| \frac{{{z}_{1}}}{k_{b1}^{2}-z_{1}^{2}} \right|-\frac{{{z}_{1}}}{k_{b1}^{2}-z_{1}^{2}}\tanh \left( \frac{\left( \frac{{{z}_{1}}}{k_{b1}^{2}-z_{1}^{2}} \right)}{{{\upsilon }_{1}}} \right) \right)\le 0.2785{{\delta }_{1}}^{*}{{\upsilon }_{1}}.\label{LM2}
\end{align}
By using Young's inequality, we have
\begin{align}
{{\sigma }_{1}}({{\tilde{\theta }}_{1}}{{\hat{\theta }}_{1}}+{{\tilde{\delta }}_{1}}{{\hat{\delta }}_{1}})\le {{\sigma }_{1}}\left( \frac{{{\delta }_{1}}^{*}+\theta _{1}^{2}}{2} \right)-{{\sigma }_{1}}\left( \frac{{{{\tilde{\delta }}}_{1}}^{2}+\tilde{\theta }_{1}^{2}}{2} \right).\label{LM21}
\end{align}
Substituting (\ref{LM2}),(\ref{LM21}) into (\ref{dNLF2}) yields
\begin{align}
{{{\dot{V}}}_{1}}\le&-\frac{{{K}_{1}}{{z}_{1}}^{2}}{k_{b1}^{2}-z_{1}^{2}}\text{+}\frac{{{z}_{1}}{{z}_{2}}}{k_{b1}^{2}-z_{1}^{2}}+\frac{e_{2}^{2}}{4}+\frac{a_{1}^{2}}{2}-{{\sigma }_{1}}\left( \frac{{{{\tilde{\delta }}}_{1}}^{2}+\tilde{\theta }_{1}^{2}}{2} \right) \nonumber \\
&+{{\mu }_{1}}({{\Gamma }_{1}})+0.2785{{\delta }_{1}}^{*}{{\upsilon }_{1}}+{{\sigma }_{1}}\left( \frac{{{\delta }_{1}}^{*}+\theta _{1}^{2}}{2} \right).\label{dNLF3}
\end{align}

To circumvent the so-called "explosion of complexity" and the problem of obtaining the analytic expression of time derivative of virtual input $\alpha_1$ in consideration of the unknown external disturbance, based on DSC technique we introduce the following low-pass filter with $w_2$ as output and $\alpha_{1}$ input.
\begin{align}
{{\tau }_{2}}{{\dot{w}}_{2}}+{{w}_{2}}={{\alpha }_{1}},\text{  }{{w}_{2}}(0)={{\alpha }_{1}}(0),\label{DSC1}
\end{align}
where $\tau_{2}$ is a designed constant of low-pass filter. Note that $e_{2}=w_{2}-\alpha_{1}$, we have
\begin{align}
{{\dot{e}}_{2}}=\frac{-{{e}_{2}}}{{{\tau }_{2}}}-{{\dot{\alpha }}_{1}}.\label{DSC2}
\end{align}
Therefore, the bound of time derivative of virtual input $\alpha_{1}$ can be written as
\begin{align}
|{{\dot{\alpha }}_{1}}\text{ }\!\!|\!\!=\left| {{{\dot{e}}}_{2}}+\frac{{{e}_{2}}}{{{\tau }_{2}}} \right|\le {{\phi }_{2}}({{\Psi }_{2}}),\label{DSC3}
\end{align}
where ${{\phi }_{2}}({{\Psi }_{2}})$ is a positive continuous function with ${{\Psi }_{2}}=[{{z}_{1}},{{z}_{2}},{{e}_{2}},{{\hat{\theta }}_{1}},{{\hat{\delta }}_{1}},{{y}_{d}},{{\dot{y}}_{d}},{{\ddot{y}}_{d}}]\in \mathbb{R}^{8}$. Combining (\ref{DSC2}),(\ref{DSC3}) yields
\begin{align}
{{e}_{2}}{{{\dot{e}}}_{2}}&=\frac{-e_{2}^{2}}{{{\tau }_{2}}}-{{e}_{2}}{{{\dot{\alpha }}}_{1}} \nonumber  \\
&\le \frac{-e_{2}^{2}}{{{\tau }_{2}}}+|{{e}_{2}}|{{\Psi }_{2}} \nonumber \\
&\le \frac{-e_{2}^{2}}{{{\tau }_{2}}}+e_{2}^{2}+\frac{\Psi _{2}^{2}}{4}. \label{DSC4}
\end{align}

\textbf{Step $i$ :}($i=2,\ldots,n-1$) Considering the time derivative of ${z_{i}}$
\begin{align}
{{\dot{z}}_{i}}={{f}_{i}}({{\bar{x}}_{i}},{{x}_{i+1}})+{{d}_{i}}(t)-{{\dot{w}}_{i}}.\label{zi}
\end{align}
From the Assumption 1, we know that $\partial {f_{i}}({{\bar{x}}_{i}},{{x}_{i+1}})/\partial x_{i+1} > g_{i0} >0, \forall ({{\bar{x}}_{i}},{{x}_{i+1}})\in \mathbb{R}^{i+1}$. Define $v_i=-\dot{w}_i$, we have the following inequality
\begin{align}
\frac{\partial {{f}_{1}}({{\bar{x}}_{i}},{{x}_{i+1}})+{{v}_{i}}}{\partial {x_{i+1}}}>{{g}_{i0}}>0
\end{align}
According ro Lemma 1, there exists smooth ideal input $x_{i+1}=\alpha _{i}^{*}(\bar{x_{i}},v_i), \forall (\bar{x_{i}},v_i)\in \mathbb{R}^{i+1}$, such that
\begin{align}
{{f}_{i}}({{\bar{x}}_{i}},\alpha _{i}^{*})+{{v}_{i}}=0.
\end{align}
By using Mean Value Theorem, there exists ${\lambda_{i}} (0<\lambda_i<1)$ satisfying
\begin{align}
{{f}_{i}}({{\bar{x}}_{i}},{{x}_{i+1}})={{f}_{i}}({{\bar{x}}_{i}},\alpha _{i}^{*})+{{g}_{i{{\lambda }_{i}}}}({{x}_{i+1}}-\alpha _{i}^{*}),\label{mthi}
\end{align}
where ${{g}_{i{{\lambda }_{i}}}}={{g}_{i}}({\bar{x}_{i}},{{x}_{(i+1){{\lambda }_{i}}}}), {{x}_{(i+1){{\lambda }_{i}}}}={{\lambda }_{i}}{{x}_{i+1}}+(1-{{\lambda }_{i}})\alpha _{i}^{*}$. Substitute (\ref{mthi}) into (\ref{zi}), we yield
\begin{align}
{{\dot{z}}_{i}}={{g}_{i{{\lambda }_{i}}}}({{x}_{i+1}}-\alpha_{i}^{*})+{{d}_{i}}(t). \label{dzi}
\end{align}

Define the Barrier Lyapunov function as follows
\begin{align}
{{V}_{zi}}=\frac{1}{2{{g}_{i{{\lambda }_{i}}}}}\ln \left( \frac{k_{bi}^{2}}{k_{bi}^{2}-z_{i}^{2}} \right),\label{BLFi}
\end{align}
where $k_{bi} \le k_{ci}-{\rho}_{i}, {\rho}_{i}$ is the bound of the DSC variable $w_i$, then the time derivative of (\ref{BLFi}) is
\begin{align}
{{\dot{V}}_{zi}}=\frac{1}{{{g}_{i{{\lambda }_{i}}}}}\left( \frac{{{z}_{i}}{{{\dot{z}}}_{i}}}{k_{bi}^{2}-z_{i}^{2}} \right)-\frac{{{{\dot{g}}}_{i{{\lambda }_{i}}}}}{2{{g}_{i{{\lambda }_{i}}}}^{2}}\ln \left( \frac{k_{bi}^{2}}{k_{bi}^{2}-z_{i}^{2}} \right).\label{dBLFi}
\end{align}
By substituting (\ref{dzi}) into (\ref{dBLFi}), we obtain
\begin{align}
{{\dot{V}}_{zi}}=\frac{{{z}_{i}}({{x}_{i+1}}-\alpha _{i}^{*})}{k_{bi}^{2}-z_{i}^{2}}+\frac{1}{{{g}_{i{{\lambda }_{i}}}}}\frac{{{z}_{i}}{{d}_{i}}(t)}{k_{b2}^{2}-z_{i}^{2}}-\frac{{{{\dot{g}}}_{i{{\lambda }_{i}}}}}{2{{g}_{i{{\lambda }_{i}}}}^{2}}\ln \left( \frac{k_{bi}^{2}}{k_{bi}^{2}-z_{i}^{2}} \right). \label{dBLFi2}
\end{align}
According to Lemma 3, $\alpha _{i}^{*}$ can be approximated by appropriate FLSs as follows
\begin{align}
\alpha _{i}^{*}({{Z}_{i}})={{W}_{i}}^{T}{{\xi }_{i}}({{Z}_{i}})+{{\varepsilon }_{i}}({{Z}_{i}}), \label{FAi1}
\end{align}
where ${{Z}_{i}}={{\left[ {\bar{x}_{i}},{{{\dot{w}}}_{i}} \right]}^{T}}\in {{R}^{i+1}}$ is the argument vector of unknown function $\alpha _{i}^{*}({\bar{x}_{i}},{{{\dot{w}}}_{i}})$. And there exists unknown constant ${{\varepsilon }_{i}}^{*}$, such that $\vert {{\varepsilon }_{i}} \vert \le {{\varepsilon }_{i}}^{*}$. Then, substituting (\ref{FAi1}) into (\ref{dBLFi2}) yields
\begin{align}
{{\dot{V}}_{zi}}=\frac{{{z}_{i}}({{z}_{i+1}}+{{e}_{i+1}}+{{\alpha }_{i}})}{k_{bi}^{2}-z_{i}^{2}}+\frac{1}{{{g}_{i{{\lambda }_{i}}}}}\frac{{{z}_{i}}{{d}_{i}}(t)}{k_{bi}^{2}-z_{i}^{2}}-\frac{{{z}_{i}}\alpha _{i}^{*}}{k_{bi}^{2}-z_{i}^{2}}-\frac{{{{\dot{g}}}_{i{{\lambda }_{i}}}}}{2{{g}_{i{{\lambda }_{i}}}}^{2}}\ln \left( \frac{k_{bi}^{2}}{k_{bi}^{2}-z_{i}^{2}} \right),\label{dBLFi3}
\end{align}
where $e_{i+1}=w_{i+1}-\alpha_{i}, x_{i+1}=z_{i+1}+e_{i+1}+\alpha_{i}$. By using Young's inequality, we have
\begin{align}
-\frac{{{z}_{i}}\alpha _{i}^{*}}{k_{bi}^{2}-z_{i}^{2}}&\le \left| \frac{{{z}_{i}}}{k_{bi}^{2}-z_{i}^{2}} \right|\left( |{{W}_{i}}^{T}\xi_{1}|+|{{\varepsilon }_{i}}| \right) \nonumber \\
&\le \frac{{{z}_{i}}^{2}{{\left\| {{W}_{i}} \right\|}^{2}}\xi_{i}^{T}\xi_{i}}{2{{a}_{\text{i}}}^{2}{{(k_{bi}^{2}-z_{i}^{2})}^{2}}}+\frac{{{a}_{\text{i}}}^{2}}{2}+\left| \frac{{{z}_{i}}}{k_{bi}^{2}-z_{i}^{2}} \right|{{\varepsilon }_{i}}^{*},\label{YIi1}
\end{align}
where $a_i>0$ is a designed constant. By substituting (\ref{YIi1}) into (\ref{dBLFi3}), we have
\begin{align}
{{{\dot{V}}}_{zi}}\le&\frac{{{z}_{i}}({{z}_{i+1}}+{{e}_{i+1}})}{k_{bi}^{2}-z_{i}^{2}}+\frac{{{z}_{i}}{{\alpha }_{i}}}{k_{bi}^{2}-z_{i}^{2}}+\frac{1}{{{g}_{i0}}}\left| \frac{{{z}_{i}}}{k_{bi}^{2}-z_{i}^{2}} \right|\left| {{d}_{i}}(t) \right|+\frac{{{z}_{i}}^{2}{{\left\| {{W}_{i}} \right\|}^{2}}\xi _{i}^{T}{{\xi }_{i}}}{2{{a}_{i}}^{2}{{(k_{bi}^{2}-z_{i}^{2})}^{2}}} \nonumber \\
&+\frac{{{a}_{\text{i}}}^{2}}{2}+\left| \frac{{{z}_{i}}}{k_{bi}^{2}-z_{i}^{2}} \right|{{\varepsilon }_{i}}^{*}-\frac{{{{\dot{g}}}_{i{{\lambda }_{i}}}}}{2{{g}_{i{{\lambda }_{i}}}}^{2}}\ln \left( \frac{k_{bi}^{2}}{k_{bi}^{2}-z_{i}^{2}} \right) \nonumber \\
\le&\frac{{{z}_{i}}({{z}_{i+1}}+{{e}_{i+1}})}{k_{bi}^{2}-z_{i}^{2}}+\frac{{{z}_{i}}{{\alpha }_{i}}}{k_{bi}^{2}-z_{i}^{2}}+\frac{{{z}_{i}}^{2}{{\theta }_{i}}\xi _{i}^{T}{{\xi }_{i}}}{{{(k_{bi}^{2}-z_{i}^{2})}^{2}}}+\left| \frac{{{z}_{i}}}{k_{bi}^{2}-z_{i}^{2}} \right|{{\delta }_{i}}^{*} \nonumber \\
&+\frac{{{a}_{\text{i}}}^{2}}{2}-\frac{{{{\dot{g}}}_{i{{\lambda }_{i}}}}}{2{{g}_{i{{\lambda }_{i}}}}^{2}}\ln \left( \frac{k_{bi}^{2}}{k_{bi}^{2}-z_{i}^{2}} \right),\label{dBLFi4}
\end{align}
where ${{\delta }_{i}}^{*}=\frac{{{d}_{iM}}}{{{g}_{i0}}}+{{\varepsilon }_{i}}^{*},\text{ }{{\theta }_{i}}=\frac{{{\left\| {{W}_{i}} \right\|}^{2}}}{2{{a}_{i}}^{2}}$. By using Young's inequality we have
\begin{align}
\frac{{{z}_{i}}{{e}_{i+1}}}{k_{bi}^{2}-z_{i}^{2}}\le \frac{{{z}_{i}}^{2}}{{{\left( k_{bi}^{2}-z_{i}^{2} \right)}^{2}}}+\frac{e_{i+1}^{2}}{4}. \label{YIi2}
\end{align}
By substituting (\ref{YIi2}) into (\ref{dBLFi4}), we obtain
\begin{align}
{{{\dot{V}}}_{zi}}\le&\frac{{{z}_{i}}{{z}_{i+1}}}{k_{bi}^{2}-z_{i}^{2}}+\frac{{{z}_{i}}{{\alpha }_{i}}}{k_{bi}^{2}-z_{i}^{2}}+\frac{{{z}_{i}}^{2}{{\theta }_{i}}\xi _{i}^{T}{{\xi }_{i}}}{{{(k_{bi}^{2}-z_{i}^{2})}^{2}}}+\frac{{{z}_{i}}^{2}}{{{(k_{bi}^{2}-z_{i}^{2})}^{2}}} \nonumber \\
&+\frac{e_{i+1}^{2}}{4}+\left| \frac{{{z}_{i}}}{k_{bi}^{2}-z_{i}^{2}} \right|{{\delta }_{i}}^{*}+\frac{a_{i}^{2}}{2}+{{\mu }_{i}}({{\Gamma }_{i}}),\label{dBLFi5}
\end{align}
where ${{\mu }_{i}}({{\Gamma }_{i}})$ is a positive continuous function, and ${{\Gamma }_{i}}=[{\bar{z}_{i}},{{y}_{d}},{{\dot{y}}_{d}}]^{T}\in \mathbb{R}^{i+2}, {\bar{z}_{i}}={{[{{z}_{1}},{{z}_{2}},\ldots ,{{z}_{i}}]}^{T}}$, satisfying $\left| \frac{{{{\dot{g}}}_{i{{\lambda }_{i}}}}}{2{{g}_{i{{\lambda }_{i}}}}^{2}}\ln \left( \frac{k_{bi}^{2}}{k_{bi}^{2}-z_{i}^{2}} \right) \right|\le {{\mu }_{i}}({{\Gamma }_{i}})$.

Then, design the virtual control law $\alpha_{i}$ and adaptation laws as follows
\begin{align}
{{\alpha }_{i}}&=-{{K}_{i}}{{z}_{i}}-\frac{{{z}_{i}}{{{\hat{\theta }}}_{i}}\xi _{i}^{T}{{\xi }_{i}}}{k_{bi}^{2}-z_{i}^{2}}-{{\hat{\delta }}_{i}}\tanh \left( \frac{\left( \frac{{{z}_{i}}}{k_{bi}^{2}-z_{i}^{2}} \right)}{{{\upsilon }_{i}}} \right)-\frac{{{z}_{i}}}{k_{bi}^{2}-z_{i}^{2}}-\frac{(k_{bi}^{2}-z_{i}^{2}){{z}_{i-1}}}{k_{b(i-1)}^{2}-z_{i-1}^{2}}\label{AAI1} \\
{{\dot{\hat{\delta }}}_{i}}&={{\gamma }_{i}}\frac{{{z}_{i}}}{k_{bi}^{2}-z_{i}^{2}}\tanh \left( \frac{\left( \frac{{{z}_{i}}}{k_{bi}^{2}-z_{i}^{2}} \right)}{{{\upsilon }_{i}}} \right)-{{\sigma }_{i}}{{\gamma }_{i}}{{\hat{\delta }}_{i}} \label{AAI2}\\
{{\dot{\hat{\theta }}}_{i}}&={{\beta }_{i}}\frac{{{z}_{i}}^{2}\xi _{i}^{T}{{\xi }_{i}}}{k_{bi}^{2}-z_{i}^{2}}-{{\sigma }_{i}}{{\beta }_{i}}{{\hat{\theta }}_{i}},\label{dALi1}
\end{align}
where ${{K}_{i}}>0,{{\beta }_{i}}>0,{{\gamma }_{i}}>0,{{\sigma }_{i}}>0,{{\upsilon }_{i}}>0$ are the designed parameters, and the term ${{\hat{\theta }}_{i}},{{\hat{\delta }}_{i}}$ are the estimates of ${{\theta }_{i}}$ and ${{\delta }_{i}^{*}}$ respectively. The term ${{\hat{\delta }}_{i}}$ is used as a robust compensator to deal with the problem of external disturbance and approximation error. Note that if we choose ${{\hat{\theta }}_{i}}(0)>0, {{\hat{\delta }}_{i}}(0)>0$, then ${{\hat{\theta }}_{i}}(t)>0, {{\hat{\delta }}_{i}}(t)>0, \forall t>0$.

Considering the following augmented Lyapunov function
\begin{align}
{{V}_{i}}=\frac{1}{2{{g}_{i{{\lambda }_{i}}}}}\ln \left( \frac{k_{bi}^{2}}{k_{bi}^{2}-z_{i}^{2}} \right)+\frac{1}{2{{\gamma }_{i}}}{{\tilde{\delta }}_{i}}^{2}+\frac{1}{2{{\beta }_{i}}}{{\tilde{\theta }}_{i}}^{2},\label{NLFi1}
\end{align}
where ${{\tilde{\delta }}_{i}}\text{=}\delta _{i}^{*}\text{-}{{\hat{\delta }}_{i}},{{\tilde{\theta }}_{i}}\text{=}{{\theta }_{i}}\text{-}{{\hat{\theta }}_{i}}$. Then the time derivative of (\ref{NLFi1}) is
\begin{align}
{{{\dot{V}}}_{i}}\le&\frac{{{z}_{i}}{{z}_{i+1}}}{k_{bi}^{2}-z_{i}^{2}}+\frac{{{z}_{i}}{{\alpha }_{i}}}{k_{bi}^{2}-z_{i}^{2}}+\frac{{{z}_{i}}^{2}{{\theta }_{i}}\xi _{i}^{T}\xi {}_{i}}{{{(k_{bi}^{2}-z_{i}^{2})}^{2}}}+\frac{{{z}_{i}}^{2}}{{{(k_{bi}^{2}-z_{i}^{2})}^{2}}}+\frac{e{{{}_{i+1}}^{2}}}{4} \nonumber \\
&+\left| \frac{{{z}_{i}}}{k_{bi}^{2}-z_{i}^{2}} \right|{{\delta }_{i}}^{*}+\frac{a_{i}^{2}}{2}+{{\mu }_{i}}({{\Gamma }_{i}})-\frac{1}{{{\gamma }_{i}}}{{{\tilde{\delta }}}_{i}}{{{\dot{\hat{\delta }}}}_{i}}-\frac{1}{{{\beta }_{i}}}{{{\tilde{\theta }}}_{i}}{{{\dot{\hat{\theta }}}}_{i}}.\label{dNLFi1}
\end{align}
By substituting (\ref{AAI1}),(\ref{AAI2}),(\ref{dALi1}) into (\ref{dNLFi1}), we obtain
\begin{align}
{{{\dot{V}}}_{i}}\le&\frac{{{z}_{i}}{{z}_{i+1}}}{k_{bi}^{2}-z_{i}^{2}}-\frac{{{z}_{i-1}}{{z}_{i}}}{k_{b\left( i-1 \right)}^{2}-z_{i-1}^{2}}-\frac{{{K}_{i}}{{z}_{i}}^{2}}{k_{bi}^{2}-z_{i}^{2}}+\frac{{{z}_{i}}^{2}{{\theta }_{i}}\xi _{i}^{T}{{\xi }_{i}}}{{{(k_{bi}^{2}-z_{i}^{2})}^{2}}}+\frac{e_{i+1}^{2}}{4}+\left| \frac{{{z}_{i}}}{k_{bi}^{2}-z_{i}^{2}} \right|{{\delta }_{i}}^{*} \nonumber \\
&+\frac{{{a}_{i}}^{2}}{2}+{{\mu }_{i}}({{\Gamma }_{i}})-\frac{1}{{{\gamma }_{i}}}{{{\tilde{\delta }}}_{i}}{{{\dot{\hat{\delta }}}}_{i}}-\frac{1}{{{\beta }_{i}}}{{{\tilde{\theta }}}_{i}}{{{\dot{\hat{\theta }}}}_{i}} \nonumber  \\
\le&\frac{{{z}_{i}}{{z}_{i+1}}}{k_{bi}^{2}-z_{i}^{2}}-\frac{{{z}_{i-1}}{{z}_{i}}}{k_{b\left( i-1 \right)}^{2}-z_{i-1}^{2}}-\frac{{{K}_{i}}{{z}_{i}}^{2}}{k_{bi}^{2}-z_{i}^{2}}+\frac{e_{i+1}^{2}}{4}-\frac{{{z}_{i}}}{k_{bi}^{2}-z_{i}^{2}}{{{\hat{\delta }}}_{i}}\tanh \left( \frac{\left( \frac{{{z}_{i}}}{k_{bi}^{2}-z_{i}^{2}} \right)}{{{\upsilon }_{i}}} \right)+\frac{{{a}_{i}}^{2}}{2} \nonumber \\
&+{{\mu }_{i}}({{\Gamma }_{i}})+{{{\tilde{\delta }}}_{i}}\left( {{\sigma }_{i}}{{{\hat{\delta }}}_{i}}-\frac{{{z}_{i}}}{k_{bi}^{2}-z_{i}^{2}}\tanh \left( \frac{\left( \frac{{{z}_{i}}}{k_{bi}^{2}-z_{i}^{2}} \right)}{{{\upsilon }_{i}}} \right) \right)+\left| \frac{{{z}_{i}}}{k_{bi}^{2}-z_{i}^{2}} \right|{{\delta }_{i}}^{*}+{{\sigma }_{i}}{{{\tilde{\theta }}}_{i}}{{{\hat{\theta }}}_{i}} \nonumber \\
\le&-\frac{{{K}_{i}}{{z}_{i}}^{2}}{k_{bi}^{2}-z_{i}^{2}}-\frac{{{z}_{i-1}}{{z}_{i}}}{k_{b\left( i-1 \right)}^{2}-z_{i-1}^{2}}\text{+}\frac{{{z}_{i}}{{z}_{i+1}}}{k_{bi}^{2}-z_{i}^{2}}+\frac{{{e}_{i+1}}^{2}}{4}+\frac{{{a}_{i}}^{2}}{2}+{{\mu }_{i}}({{\Gamma }_{i}}) \nonumber  \\
&+{{\delta }_{i}}^{*}\left( \left| \frac{{{z}_{i}}}{k_{bi}^{2}-z_{i}^{2}} \right|-\frac{{{z}_{i}}}{k_{b1}^{i+1}-z_{i}^{2}}\tanh \left( \frac{\left( \frac{{{z}_{i}}}{k_{bi}^{2}-z_{i}^{2}} \right)}{{{\upsilon }_{i}}} \right) \right) \nonumber \\
&+{{\sigma }_{i}}({{{\tilde{\theta }}}_{i}}{{{\hat{\theta }}}_{i}}+{{{\tilde{\delta }}}_{i}}{{{\hat{\delta }}}_{i}}).\label{dNLFi2}
\end{align}
According to Lemma 2, we obtain
\begin{align}
{{\delta }_{i}}^{*}\left( \left| \frac{{{z}_{i}}}{k_{bi}^{2}-z_{i}^{2}} \right|-\frac{{{z}_{i}}}{k_{bi}^{2}-z_{i}^{2}}\tanh \left( \frac{\left( \frac{{{z}_{i}}}{k_{bi}^{2}-z_{i}^{2}} \right)}{{{\upsilon }_{i}}} \right) \right)\le 0.2785{{\delta }_{i}}^{*}{{\upsilon }_{i}}.\label{LMi2}
\end{align}
By using Young's inequality, we have
\begin{align}
{{\sigma }_{i}}({{\tilde{\theta }}_{i}}{{\hat{\theta }}_{i}}+{{\tilde{\delta }}_{i}}{{\hat{\delta }}_{i}})\le {{\sigma }_{i}}\left(\frac{{{\delta }_{i}}^{*}+\theta _{i}^{2}}{2}\right)-{{\sigma }_{i}}\left(\frac{{{{\tilde{\delta }}}_{i}}^{2}+\tilde{\theta }_{i}^{2}}{2}\right).\label{LMi21}
\end{align}
Substituting (\ref{LMi2}),(\ref{LMi21}) into (\ref{dNLFi2}) yields
\begin{align}
{{\dot{V}}_{i}}\le&-\frac{{{K}_{i}}{{z}_{i}}^{2}}{k_{bi}^{2}-z_{i}^{2}}-\frac{{{z}_{i-1}}{{z}_{i}}}{k_{b\left( i-1 \right)}^{2}-z_{i-1}^{2}}+\frac{{{z}_{i}}{{z}_{i+1}}}{k_{bi}^{2}-z_{i}^{2}}+\frac{e_{i+1}^{2}}{4}+\frac{a_{i}^{2}}{2}-{{\sigma }_{i}}\left(\frac{{{{\tilde{\delta }}}_{i}}^{2}+\tilde{\theta }_{i}^{2}}{2}\right) \nonumber \\
&+{{\mu }_{i}}({{\Gamma }_{i}})+0.2785{{\delta }_{i}}^{*}{{\upsilon }_{i}}+{{\sigma }_{i}}\left(\frac{{{\delta }_{i}}^{*}+\theta _{i}^{2}}{2}\right).\label{dNLFi3}
\end{align}

Based on DSC technique, we introduce the following low-pass filter with $w_{i+1}$ as output and $\alpha_{i}$ input.
\begin{align}
{{\tau }_{i+1}}{{\dot{w}}_{i+1}}+{{w}_{i+1}}={{\alpha }_{i}},\text{  }{{w}_{i+1}}(0)={{\alpha }_{i}}(0),\label{DSCi1}
\end{align}
where $\tau_{i+1}$ is a designed constant of low-pass filter. Note that $e_{i+1}=w_{i+1}-\alpha_{i}$, we have
\begin{align}
{{\dot{e}}_{i+1}}=\frac{-{{e}_{i+1}}}{{{\tau }_{i+1}}}-{{\dot{\alpha }}_{i}}.\label{DSCi2}
\end{align}
Therefore, the bound of time derivative of virtual input $\alpha_{i}$ can be written as
\begin{align}
|{{\dot{\alpha }}_{i}}\text{ }\!\!|\!\!=\left| {{{\dot{e}}}_{i+1}}+\frac{{{e}_{i+1}}}{{{\tau }_{i+1}}} \right|\le {{\phi }_{i+1}}({{\Psi }_{i+1}}),\label{DSCi3}
\end{align}
where ${{\phi }_{i+1}}({{\Psi }_{i+1}})$ is a positive continuous function with \\
${{\Psi }_{i+1}}=[{\bar{z}_{i+1}},{\bar{e}_{i+1}},{{\bar{\hat{\theta }}}_{i}},{{\bar{\hat{\delta }}}_{i}},{{y}_{d}},{{\dot{y}}_{d}},{{\ddot{y}}_{d}}]\in \mathbb{R}^{4i+3}$. Combining (\ref{DSCi2}),(\ref{DSCi3}) yields
\begin{align}
{{e}_{i+1}}{{{\dot{e}}}_{i+1}}&=\frac{-e_{i+1}^{2}}{{{\tau }_{i+1}}}-{{e}_{i+1}}{{{\dot{\alpha }}}_{i}} \nonumber  \\
&\le \frac{-e_{i+1}^{2}}{{{\tau }_{i+1}}}+|{{e}_{i+1}}|{{\Psi }_{i+1}} \nonumber \\
&\le \frac{-e_{i+1}^{2}}{{{\tau }_{i+1}}}+e_{i+1}^{2}+\frac{\Psi _{i+1}^{2}}{4}. \label{DSCi4}
\end{align}

\textbf{Step $n$ :} Considering the time derivative of ${z_{n}}$
\begin{align}
{{\dot{z}}_{n}}={{f}_{n}}({{\bar{x}}_{n}},\chi (t)-u(u))+{{d}_{n}}(t)-{{\dot{w}}_{n}},\label{zn}
\end{align}
similar to the first $n-1$ steps, (\ref{zn}) can be rewritten as
\begin{align}
{{\dot{z}}_{n}}={{g}_{n{{\lambda }_{n}}}}\left( \chi (t)-u(t)-\alpha _{n}^{*} \right)+{{d}_{n}}\text{(t) },
\end{align}
where ${{g}_{n{{\lambda }_{n}}}}={{g}_{n}}({\bar{x}_{n}},{{x}_{n+1{{\lambda }_{n}}}}), {{x}_{n+1{{\lambda }_{n}}}}={{\lambda }_{n}}(\chi (t)-u(t))+(1-{{\lambda }_{i}})\alpha _{n}^{*}$.

Define the following Lyapunov function
\begin{align}
{{V}_{n}}=\frac{1}{2{{g}_{n{{\lambda }_{n}}}}}\ln \left( \frac{k_{bn}^{2}}{k_{bn}^{2}-z_{n}^{2}} \right)\text{ +}\frac{1}{2{{\gamma }_{n}}}{{\tilde{\delta }}_{n}}^{2}+\frac{1}{2{{\beta }_{n}}}{{\tilde{\theta }}_{n}}^{2},\label{LFN}
\end{align}
where $k_{bn} \le k_{cn}-{\rho}_{n}, {\rho}_{n}$ is the bound of the DSC variable $w_n$, ${{\tilde{\theta }}_{n}}={{\theta }_{n}}-{{\hat{\theta }}_{n}},{{\tilde{\delta }}_{n}}=\delta _{n}^{*}-{{\hat{\delta }}_{n}}$, and $\gamma_{n}>0,\beta_{n}>0$ are designed contants.
Furthermore, the time derivative of (\ref{LFN}) is
\begin{align}
{{{\dot{V}}}_{n}}&\le \frac{{{z}_{n}}{{z}_{n+1}}}{k_{bn}^{2}-z_{n}^{2}}+\frac{{{z}_{n}}{{\alpha }_{n}}}{k_{bn}^{2}-z_{n}^{2}}+\frac{{{z}_{n}}^{2}{{\theta }_{n}}\xi _{1}^{T}{{\xi }_{1}}}{{{(k_{bn}^{2}-z_{n}^{2})}^{2}}}+\left| \frac{{{z}_{1}}}{k_{b1}^{2}-z_{1}^{2}} \right|{{\delta }_{1}}^{*}+\frac{a_{n}^{2}}{2} \nonumber \\
&+{{\mu }_{n}}({{\Gamma }_{n}})-\frac{1}{{{\gamma }_{n}}}{{{\tilde{\delta }}}_{n}}{{{\dot{\hat{\delta }}}}_{n}}-\frac{1}{{{\beta }_{n}}}{{{\tilde{\theta }}}_{n}}{{{\dot{\hat{\theta }}}}_{n}},\label{dLFN1}
\end{align}
where ${{z}_{n+1}}=\chi (t)-u(t)-{{\alpha }_{n}},\delta _{n}^{*}=\frac{{{d}_{nM}}}{{{g}_{n0}}}+\varepsilon _{n}^{*},{{\theta }_{n}}=\frac{{{\left\| {{W}_{n}} \right\|}^{2}}}{2a_{n}^{2}},{{a}_{n}}>0$ is designed constant. ${{\mu }_{n}}({{\Gamma }_{n}})$ is a positive continuous function, and ${{\Gamma }_{n}}=[{\bar{z}_{n}},{{y}_{d}},{{\dot{y}}_{d}}]^{T}\in \mathbb{R}^{n+2}, {\bar{z}_{n}}={{[{{z}_{1}},{{z}_{2}},\ldots ,{{z}_{n}}]}^{T}}$, satisfying $\left| \frac{{{{\dot{g}}}_{n{{\lambda }_{n}}}}}{2{{g}_{n{{\lambda }_{n}}}}^{2}}\ln \left( \frac{k_{bn}^{2}}{k_{bn}^{2}-z_{n}^{2}} \right) \right|\le {{\mu }_{n}}({{\Gamma }_{n}})$.

Design virtual control law $\alpha_{n}$ and adaptation laws as follows
\begin{align}
{{\alpha }_{n}}&=-{{K}_{n}}{{z}_{n}}-\frac{{{z}_{n}}{{{\hat{\theta }}}_{n}}\xi _{n}^{T}{{\xi }_{n}}}{k_{bn}^{2}-z_{n}^{2}}-{{\hat{\delta }}_{n}}\tanh \left( \frac{\left( \frac{{{z}_{n}}}{k_{bn}^{2}-z_{n}^{2}} \right)}{{{\upsilon }_{n}}} \right)-\frac{(k_{bn}^{2}-z_{n}^{2}){{z}_{n-1}}}{k_{b(n-1)}^{2}-z_{n-1}^{2}},\label{AAN1} \\
{{\dot{\hat{\delta }}}_{n}}&={{\gamma }_{n}}\frac{{{z}_{n}}}{k_{bn}^{2}-z_{n}^{2}}\tanh \left( \frac{\left( \frac{{{z}_{n}}}{k_{bn}^{2}-z_{n}^{2}} \right)}{{{\upsilon }_{n}}} \right)-{{\sigma }_{n}}{{\gamma }_{n}}{{\hat{\delta }}_{n}},\label{VIn1} \\
{{\dot{\hat{\theta }}}_{n}}&={{\beta }_{n}}\frac{{{z}_{n}}^{2}\xi _{n}^{T}{{\xi }_{n}}}{k_{bn}^{2}-z_{n}^{2}}-{{\sigma }_{n}}{{\beta }_{n}}{{\hat{\theta }}_{n}}, \label{dAFN11}
\end{align}
where ${{K}_{i}}>0,{{\sigma }_{i}}>0,{{\upsilon }_{i}}>0$ are the designed parameters.
Substituting (\ref{AAN1})-(\ref{dAFN11}) into (\ref{dLFN1}) yields
\begin{align}
{{{\dot{V}}}_{n}}\le&-\frac{{{K}_{n}}z_{n}^{2}}{k_{bn}^{2}-z_{n}^{2}}\text{+}\frac{{{z}_{n}}{{z}_{n+1}}}{k_{bn}^{2}-z_{n}^{2}}-\frac{{{z}_{n-1}}{{z}_{n}}}{k_{b\left( n-1 \right)}^{2}-z_{n-1}^{2}}+\frac{e_{n+1}^{2}}{4}-{{\sigma }_{n}}\left( \frac{{{{\tilde{\delta }}}_{n}}^{2}+\tilde{\theta }_{n}^{2}}{2} \right) \nonumber \\
&+{{\mu }_{n}}({{\Gamma }_{n}})+\frac{a_{n}^{2}}{2}+0.2785{{\delta }_{n}}^{*}{{\upsilon }_{n}}\text{+}{{\sigma }_{n}}\left( \frac{{{\delta }_{n}}^{*}+\theta _{n}^{2}}{2} \right).\label{dLFF}
\end{align}

Based on DSC technique, we have
\begin{align}
{{e}_{n+1}}{{\dot{e}}_{n+1}}=\frac{-e_{n+1}^{2}}{{{\tau }_{n+1}}}-{{e}_{n+1}}{{\dot{\alpha }}_{n}}\le \frac{-e_{n+1}^{2}}{{{\tau }_{n+1}}}&+|{{e}_{n+1}}|{{\phi }_{n+1}}\le \frac{-e_{n+1}^{2}}{{{\tau }_{n+1}}}+e_{n+1}^{2}+\frac{\phi _{n+1}^{2}}{4}, \\
{{\tau }_{n+1}}{{\dot{w}}_{n+1}}+{{w}_{n+1}}={{\alpha }_{n}}&,\text{  }{{w}_{n+1}}(0)={{\alpha }_{n}}(0),
\end{align}
where ${{e}_{n+1}}={{w}_{n+1}}-{{\alpha }_{n}}$, and $\tau_{n+1}$ is the designed parameter of filter. ${{\phi }_{n+1}}({{\Psi }_{n+1}})$ is a positive continuous function, and ${{\Psi }_{n+1}}=[{{\bar{z}}_{n+1}},{{\bar{e}}_{n+1}},{{\bar{\hat{\theta }}}_{n}},{{\bar{\hat{\delta }}}_{n}},{{y}_{d}},{{\dot{y}}_{d}},{{\ddot{y}}_{d}}]\in {{\mathbb{R}}^{4n+5}}$.

\textbf{Step $n+1$ :} Choosing the quadratic function $V_{n+1}$ as $V_{n+1}={1}/{2}z_{n+1}^{2}$, we have
\begin{align}
{{\dot{V}}_{n+1}}={{z}_{n+1}}\left( \dot{\chi }(t)-\dot{u}(t)-{{{\dot{w}}}_{n+1}} \right).\label{dLNP1}
\end{align}
Substituting transformed system (\ref{Xn 3}) into (\ref{dLNP1}) yields
\begin{align}
{{\dot{V}}_{n+1}}={{z}_{n+1}}\left( -\lambda \chi (t)+\left( 2\lambda +\kappa  \right)u(t)-v-{{{\dot{w}}}_{n+1}} \right),
\end{align}
choosing $v={{K}_{n+1}}{{z}_{n+1}}-\lambda \chi (t)+\left( 2\lambda +\kappa  \right)u(t)+{{e}_{n+1}}/{{\tau }_{i+1}}$, where $K_{n+1}>0$ is designed constant. We have
\begin{align}
{{\dot{V}}_{n+1}}=-{{K}_{n+1}}z_{n+1}^{2}.\label{Final}
\end{align}

\section{Stability Analysis and Feasibility Check}
In this section, we will show that the semi-globally ultimately uniform boundedness of all signals in the closed-loop system can be guaranteed, and the tracking error can be bounded in a designed small neighbor of zero, while full states constraints of the pure-feedback system (\ref{pf plant}) remain within the predefined constraint sets.

Choosing the following Lyapunov function
\begin{align}
V=\sum\limits_{i=1}^{n+1}{{{V}_{i}}}+\frac{1}{2}\sum\limits_{i=2}^{n+1}{{{e}_{i}}^{2}}.\label{SLF1}
\end{align}

Note that $\ln \left( \frac{k_{bi}^{2}}{k_{bi}^{2}-z_{i}^{2}} \right)\le \frac{z_{i}^{2}}{k_{bi}^{2}-z_{i}^{2}}$ in the interval $|{{z}_{i}}|<{{k}_{bi}}$, combine (\ref{dNLF3}),(\ref{DSC4}),(\ref{dNLFi3}), \\
(\ref{DSCi4}),(\ref{Final}), therefore the time derivative of (\ref{SLF1}) is
\begin{align}
\dot{V}\le&-{{K}_{n+1}}z_{n+1}^{2}-\sum\limits_{i=1}^{n}{\frac{{{K}_{i}}{{z}_{i}}^{2}}{k_{bi}^{2}-z_{i}^{2}}}\text{+}\sum\limits_{i=2}^{n+1}{\frac{{{e}_{i}}^{2}}{4}}+\sum\limits_{i=1}^{n}{\left( -\frac{{{e}_{i+1}}^{2}}{{{\tau }_{i+1}}}+{{e}_{i+1}}^{2}+\frac{{{\phi }_{i}}{{({{\Psi }_{i}})}^{2}}}{4} \right)} \nonumber \\
&-\sum\limits_{i=1}^{n}{\left( {{\sigma }_{i}}\left(\frac{{{{\tilde{\delta }}}_{i}}^{2}+\tilde{\theta }_{i}^{2}}{2}\right)-{{\mu }_{i}}({{\Gamma }_{i}})-0.2785{{\delta }_{i}}^{*}{{\upsilon }_{i}}\text{-}{{\sigma }_{i}}\left(\frac{{{\delta }_{i}}^{*}+\theta _{i}^{2}}{2}\right) \right)}+\sum\limits_{i=1}^{n}{\frac{{{a}_{i}}^{2}}{2}} \nonumber \\
=&-{{K}_{n+1}}z_{n+1}^{2}-\sum\limits_{i=1}^{n}{\frac{{{K}_{i}}{{z}_{i}}^{2}}{k_{bi}^{2}-z_{i}^{2}}}-\sum\limits_{i=1}^{n}{{{\sigma }_{i}}\left(\frac{{{{\tilde{\delta }}}_{i}}^{2}+\tilde{\theta }_{i}^{2}}{2}\right)}+\sum\limits_{i=2}^{n+1}{\left( \left(\frac{5}{4}-\frac{1}{{{\tau }_{i}}}\right){{e}_{i}}^{2} \right)} \nonumber \\
&+\sum\limits_{i=1}^{n}{\left( \frac{{{a}_{i}}^{2}}{2}+{{\mu }_{i}}({{\Gamma }_{i}})+0.2785{{\delta }_{i}}^{*}{{\upsilon }_{i}}+{{\sigma }_{i}}\left(\frac{{{\delta }_{i}}^{*}+\theta _{i}^{2}}{2}\right) \right)}+\sum\limits_{i=2}^{n+1}{\frac{{{\phi }_{i}}{{({{\Psi }_{i}})}^{2}}}{4}}.\label{dSLF}
\end{align}
Define compact sets as follows
\begin{equation}
{ \left\{ \begin{array}{l}
{{\Omega }_{1}}=\{{{[{{z}_{1}},{{{\hat{\theta }}}_{1}},{{{\hat{\delta }}}_{1}}]}^{T}}:{{V}_{1}}\le \varpi \}\subset {{\mathbb{R}}^{3}} \\
{{\Omega }_{i}}=\{{{[{{{\bar{z}}}_{i}}^{T},{{{\bar{e}}}_{i}}^{T},{{{\bar{\hat{\theta }}}}_{i}}^{T},{{{\bar{\hat{\delta }}}}_{i}}^{T}]}^{T}}:\sum\limits_{j=1}^{i}{{{V}_{i}}}+\sum\limits_{j=2}^{i}{{{e}_{j}}^2/2}\le \varpi \}\subset {{\mathbb{R}}^{4i-1}},i=2,\ldots ,n
\end{array} \right.}
\end{equation}
where $\varpi$ is a designed constant. Since ${{\Omega }_{0}}\times {{\Omega }_{i}}\subset \mathbb{R}^{4i+2}$ and ${{\Omega }_{0}}\times {{\Omega }_{i+1}}\subset \mathbb{R}^{4i+6}$ are compact sets, thus it's easy to see that $\mu_{i}(\Gamma_{i})$ has maximum $H_i$ on ${{\Omega }_{0}}\times {{\Omega }_{i}}$ and ${{\phi }_{i+1}}\left( {{\psi }_{i+1}} \right)$ has maximum ${{M}_{i+1}}$ on ${{\Omega }_{0}}\times {{\Omega }_{i+1}}$.

For any positive constant $c_1,c_2$, select parameters of adaptation control law as $K_{i}={c_1}/{2g_{i0}},{1}/{\tau_{i+1}}={5}/{4}+c_2$ and $\eta ={{\min }_{i=1,\ldots ,n}}\{{{c}_{1}},{{c}_{2}},{{\sigma }_{i}}{{\gamma }_{i}},{{\sigma }_{i}}{{\beta }_{i}},{{K}_{n+1}}\}$, where $\eta$ is a designed constant. Then, we have
\begin{align}
\dot{V}\le-\eta V+ D.\label{ff}
\end{align}
where $D=\sum\limits_{i=1}^{n}{\left( \frac{a_{i}^{2}}{2}+{{\mu }_{i}}({{\Gamma }_{i}})+0.2785{{\delta }_{i}}^{*}{{\upsilon }_{i}}\text{+}{{\sigma }_{i}}\left( \frac{{{\delta }_{i}}^{*}+\theta _{i}^{2}}{2} \right) \right)}+\sum\limits_{i=2}^{n}{\frac{{{\phi }_{i}}{{({{\Psi }_{i}})}^{2}}}{4}}$.
If $V(0) \le \varpi$, thus $\mu_{i}(\Gamma_{i}) \le H_{i},{{{\phi }_{i+1}}\left( {{\psi }_{i+1}} \right)}^{2} \le M_{i+1}^{2}$. Then
\begin{align}
D(0)\le\sum\limits_{i=1}^{n}{\left( \frac{a_{i}^{2}}{2}+{H_{i}}+0.2785{{\delta }_{i}}^{*}{{\upsilon }_{i}}+{{\sigma }_{i}}\left( \frac{{{\delta }_{i}}^{*}+\theta _{i}^{2}}{2} \right) \right)}+\sum\limits_{i=2}^{n}{\frac{M_{i+1}^{2}}{4}}=\omega,
\end{align}
select $\eta > \frac{\omega}{\varpi}$ to guarantee $\dot{V}(0) \le 0$, thus $V(t) \le \varpi, \forall t>0$.

Multiplying (\ref{ff}) by $e^{\eta t}$ on both sides and integrating, we have
\begin{align}
V(t)\le \left( V(0)-\frac{D}{\eta } \right){{e}^{-\eta t}}+\frac{D}{\eta }\le V(0){{e}^{-\eta t}}+\frac{\omega }{\eta }
\end{align}
considering $V_{i} \le V,i=1,\ldots ,n+1$, we obtain
\begin{equation}
{ \left\{ \begin{array}{l}
|{{z}_{i}}|\le {{k}_{bi}}\sqrt{1-{{e}^{-2V(0){{e}^{-\eta t}}-2\frac{\omega }{\eta }}}} \\
|{{\tilde{\theta }}_{i}}|\le \sqrt{2{{\beta }_{i}}V(0){{e}^{-\eta t}}+2{{\beta }_{i}}\frac{\omega }{\eta }} \\
|{{\tilde{\delta }}_{i}}|\le \sqrt{2{{\gamma }_{i}}V(0){{e}^{-\eta t}}+2{{\gamma }_{i}}\frac{\omega }{\eta }} \\
|{{e}_{i}}|\le \sqrt{2V(0){{e}^{-\eta t}}+2\frac{\omega }{\eta }}
\end{array} \right.}
\end{equation}
therefore, all of the signals in the closed-loop system are semiglobally ultimately uniform bounded and the bound of tracking error can be guaranteed an arbitrarily small neighbor of zero by choosing appropriate designed constant of adaptive controllers, while the full states constraints remain within predefined sets, since $\vert x_{i} \vert \le \vert z_{i} \vert+\vert w_{i} \vert < k_{bi}+\rho_{i} \le k_{ci}$.

The above control design and analysis are based on the following prerequisites:\\

a. There exists a group of positive constants $k_{ci},i=1,\ldots ,n$, satisfying that ${{k}_{bi}}\le {{k}_{ci}}-{{\rho }_{i}},i=1,\ldots ,n$ in the set ${{\Omega }}=\{{{\bar{z}}_{n}}\in {{\mathbb{R}}^{n}},{{\bar{e}}_{n}}\in {{\mathbb{R}}^{n-1}}, {{\bar{y}}_{d}}\in {{\mathbb{R}}^{3}}:|{{z}_{i}}|\le k_{bi}\sqrt{1-e^{-2V(t)}},|{{e}_{j}}|\le \sqrt{2V(t)}, |{{y}_{d}}|\le {{A}_{0}},|y_{d}^{(i)}|\le {{A}_{i}},i=1,2,j=2,\ldots,n+1\}$.\\

b. The initial states satisfy $\left| {{z}_{i}}(0) \right|<{{k}_{ci}},i=1,\ldots ,n$. 

It's necessary to check the feasibility of the above prerequisites as a priori. To make the convergence faster and loosen the constraints of $\left| {{z}_{i}}(0) \right|$, denote the solution $\varsigma ={{\left[ {{K}_{1}},\ldots ,{{K}_{n-1}},{{k}_{b2}},\ldots ,{{k}_{bn}} \right]}^{T}}$ for the following static nonlinear constrained programming
\begin{align}
\underset{{{K}_{1}},\ldots ,{{K}_{n-1}},{{k}_{b2}},\ldots ,{{k}_{bn}}>0}{\mathop{\max }}\,N\left( \varsigma  \right)=\sum\limits_{j=1}^{n-1}{{{K}_{j}}}+\sum\limits_{j=2}^{n}{{{k}_{bj}}}
\end{align}
subject to the constraints
\begin{align}
{{k}_{ci}}&>\rho_{i}(\varsigma)+{{k}_{bi}} \nonumber \\
{{k}_{bi}}&>\left| {{x}_{i}}\left( 0 \right)-{{\alpha }_{i-1}}\left( 0 \right) \right|,i=2,\ldots ,n
\end{align}
where $\rho_{i}(\varsigma)=\underset{({{\bar{z}}_{n}}, {{\bar{e}}_{n}}, {{\bar{y}}_{d}})\in {{\Omega }}}{\mathop{\sup }}\,\left| {w_{i}}\left( \kappa  \right) \right|,i=2,\ldots ,n$
\section{Simulation Illustration}
In this section, simulation studies are provided to demonstrated the effectiveness of proposed control method.

Consider the following pure-feedback nonlinear system with input delay
\begin{equation}
{\left\{\begin{array}{l}
{{{\dot{x}}}_{1}}=0.2{{x}_{1}}+10{{x}_{2}}\\
{{{\dot{x}}}_{2}}=0.6{{e}^{-x_{1}^{4}x_{2}^{2}}}+\left( 10+0.5{{e}^{-x_{2}^{2}}} \right)u(t-\tau )+0.4\sin (u(t-\tau) ) \\
y=x_{1}.
\end{array} \right.}
\end{equation}
where $y$ and $u$ are output and input of the system respectively, ${{x}_{1}},{{x}_{2}}$ are states of the system. $\tau$ is chosen as $0.01s$ and the desired output trajectory is given as ${{y}_{d}}=1.5\sin (t)+\cos (t)$. The states $x_{1},x_{2}$ are constrained by $\left| {{x}_{1}} \right|\le {{k}_{c1}}=3.8,\left| {{x}_{2}} \right|\le {{k}_{c2}}=6$

The adaptive fuzzy controllers and adaptation laws are given as follows
\begin{align}
{{\alpha }_{1}}=-{{K}_{1}}{{z}_{1}}-\frac{{{z}_{1}}{{{\hat{\theta }}}_{1}}\xi _{1}^{T}({{Z}_{1}}){{\xi }_{1}}({{Z}_{1}})}{k_{b1}^{2}-z_{1}^{2}}-{{\hat{\delta }}_{1}}\tanh \left( \frac{\left( \frac{{{z}_{1}}}{k_{b1}^{2}-z_{1}^{2}} \right)}{{{\upsilon }_{1}}} \right)-\frac{{{z}_{1}}}{k_{b1}^{2}-z_{1}^{2}}
\end{align}
\begin{align}
{{\alpha }_{2}}=-{{K}_{2}}{{z}_{2}}-\frac{{{z}_{2}}{{{\hat{\theta }}}_{2}}\xi _{2}^{T}({{Z}_{2}}){{\xi }_{2}}({{Z}_{2}})}{k_{b2}^{2}-z_{2}^{2}}-{{\hat{\delta }}_{2}}\tanh \left( \frac{\left( \frac{{{z}_{1}}}{k_{b1}^{2}-z_{1}^{2}} \right)}{{{\upsilon }_{1}}} \right)-\frac{{{z}_{1}}(k_{b2}^{2}-z_{2}^{2})}{k_{b1}^{2}-z_{1}^{2}}
\end{align}
\begin{align}
v={{K}_{n+1}}{{z}_{n+1}}-\lambda \chi (t)+\left( 2\lambda +\kappa  \right)u(t)+{{e}_{3}}/{{{\tau }_{3}}}\;
\end{align}
\begin{equation}
{ \left\{ \begin{array}{l}
{{{\dot{\hat{\delta }}}}_{1}}={{\gamma }_{1}}\frac{{{z}_{1}}}{k_{b1}^{2}-z_{1}^{2}}\tanh \left( \frac{\left( \frac{{{z}_{1}}}{k_{b1}^{2}-z_{1}^{2}} \right)}{{{\upsilon }_{1}}} \right)-{{\sigma }_{1}}{{\gamma }_{1}}{{{\hat{\delta }}}_{1}} \\
{{{\dot{\hat{\theta }}}}_{1}}={{\beta }_{1}}\frac{{{z}_{1}}\xi _{1}^{T}({{Z}_{1}}){{\xi }_{1}}({{Z}_{1}})}{k_{b1}^{2}-z_{1}^{2}}-{{\sigma }_{1}}{{\beta }_{1}}{{{\hat{\theta }}}_{1}} \\
{{{\dot{\hat{\delta }}}}_{2}}={{\gamma }_{2}}\frac{{{z}_{2}}}{k_{b2}^{2}-z_{2}^{2}}\tanh \left( \frac{\left( \frac{{{z}_{1}}}{k_{b1}^{2}-z_{1}^{2}} \right)}{{{\upsilon }_{1}}} \right)-{{\sigma }_{2}}{{\gamma }_{2}}{{{\hat{\delta }}}_{2}} \\
{{{\dot{\hat{\theta }}}}_{2}}={{\beta }_{2}}\frac{{{z}_{2}}\xi _{2}^{T}({{Z}_{2}}){{\xi }_{2}}({{Z}_{2}})}{k_{b2}^{2}-z_{2}^{2}}-{{\sigma }_{2}}{{\beta }_{2}}{{{\hat{\theta }}}_{2}} \\
\end{array} \right.}
\end{equation}
where the input $u(t)$ is generated as the output of a low-pass filter $\dot{u}=-\kappa u+v$. $Z_{1}={{[{{x}_{1}},{{\dot{y}}_{d}}]}^{T}}$ and $Z_{2}={{[{{x}_{1}},{{x}_{2}},{{\dot{w}}_{2}}]}^{T}}$. By the feasibility check, the designed constants are selected through function fseminf.m in Matlab as $K_{1}=4.9,K_{2}=10.2,K_{3}=20,\kappa=0.0001,\lambda=100,{{\beta }_{1}}=10,{{\beta }_{2}}=10,{{\sigma }_{1}}=10,{{\sigma }_{2}}=8,{{\gamma }_{1}}=10,{{\gamma }_{2}}=10,{{\upsilon }_{1}}={{\upsilon }_{2}}=0.1,{{k}_{b1}}=2,{{k}_{b2}}=5$.

\begin{figure}[htbp]
\begin{centering}
\includegraphics[scale=0.6]{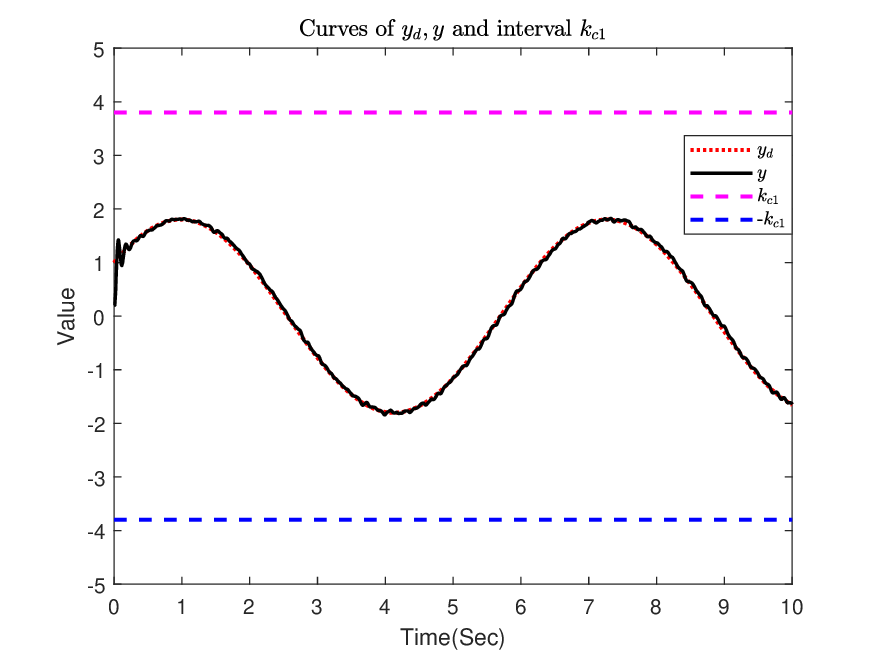}
\par\end{centering}
\caption{\label{P1}Curves of $y_{d},y$ and interval $k_{c1}$}
\end{figure}

\begin{figure}[htbp]
\begin{centering}
\includegraphics[scale=0.6]{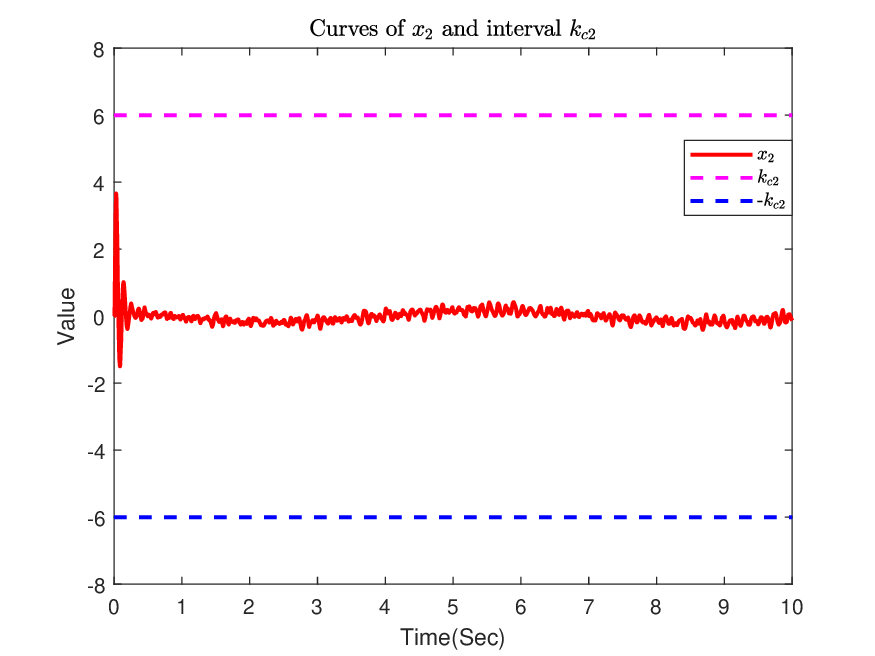}
\par\end{centering}
\caption{\label{P2}Curves of $x_{2}$ and interval $k_{c2}$}
\end{figure}

\begin{figure}[htbp]
\begin{centering}
\includegraphics[scale=0.5]{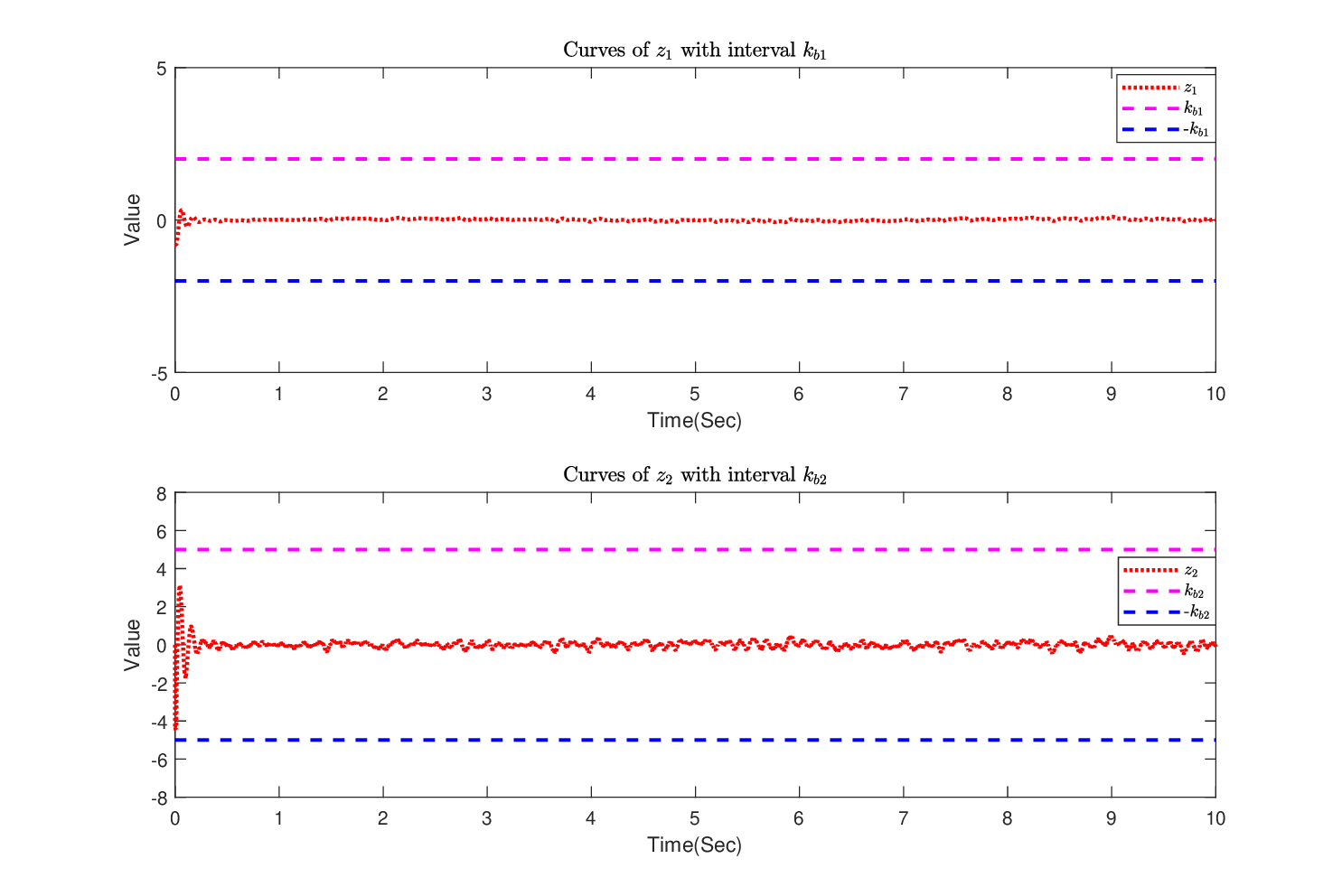}
\par\end{centering}
\caption{\label{P3}Curves of $z_{1},z_{2}$ and intervals $k_{b1},k_{b2}$}
\end{figure}

\begin{figure}[htbp]
\begin{centering}
\includegraphics[scale=0.6]{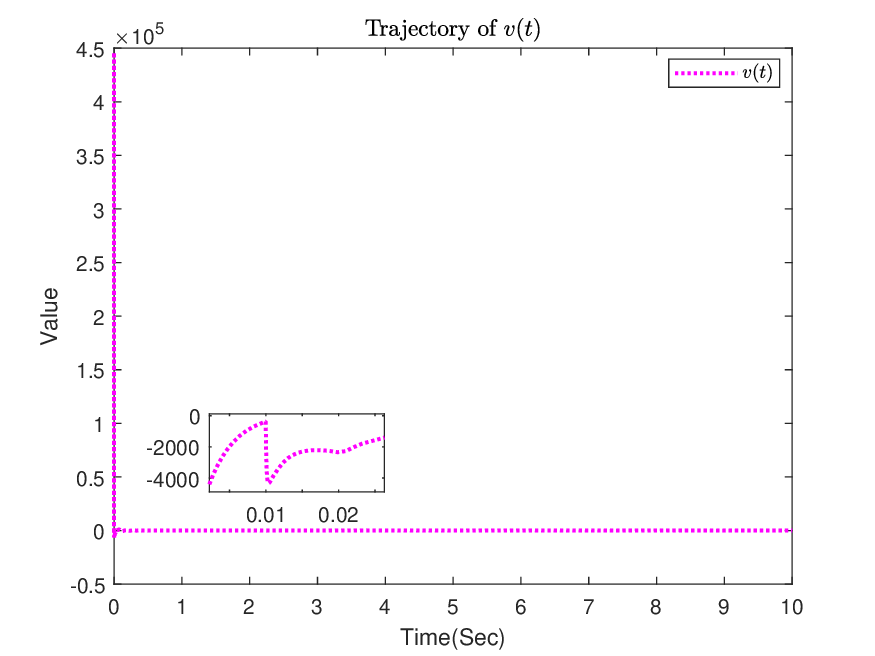}
\par\end{centering}
\caption{\label{P4}Trajectory of $v(t)$}
\end{figure}

\begin{figure}[htbp]
\begin{centering}
\includegraphics[scale=0.6]{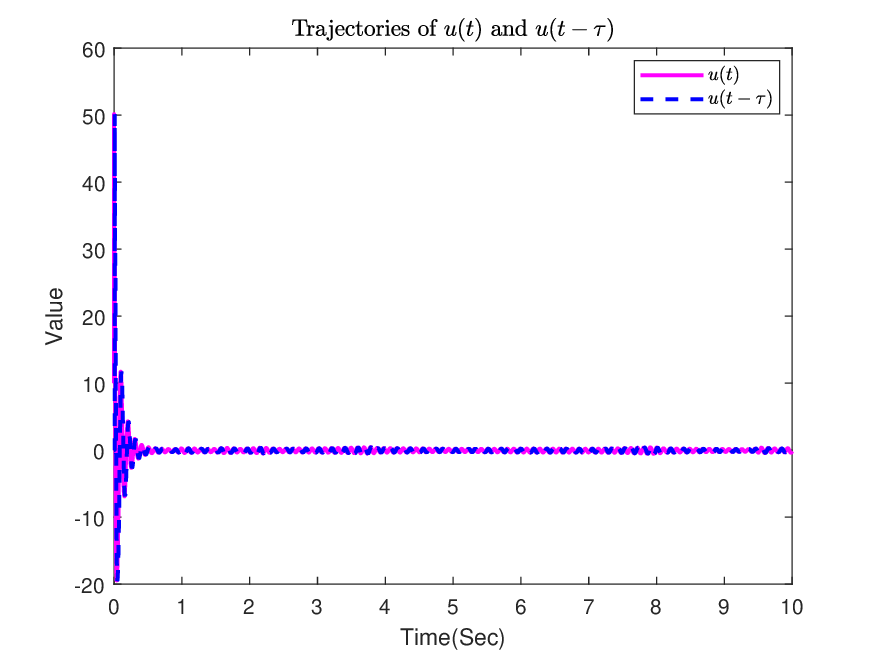}
\par\end{centering}
\caption{\label{P5}Trajectories of $u(t)$ and $u(t-{\tau})$}
\end{figure}

\begin{figure}[htbp]
\begin{centering}
\includegraphics[scale=0.6]{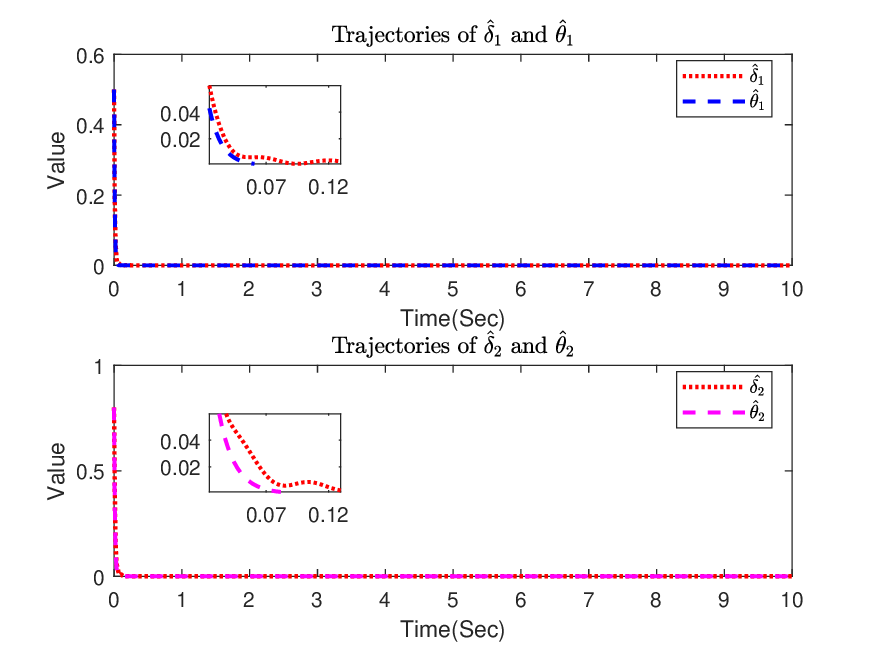}
\par\end{centering}
\caption{\label{P6}Trajectories of $\hat\delta_{1},\hat\theta_{1}$ and $\hat\delta_{2},\hat\theta_{2}$}
\end{figure}

The simulation results are shown in Figs. \ref{P1}-\ref{P6}. Fig. \ref{P1} depicts the curves of desired output $y_{d}$, the real output $y(t)$ and the state constraint interval $k_{c1}$. Fig. \ref{P2} shows the state $x_{2}$ with its constraint bound $k_{c2}$. The coordinates $z_{1},z_{2}$ are bounded in the predefined intervals $k_{b1}$ and $k_{b2}$ respectively in Fig. \ref{P3}. Fig. \ref{P4} shows the trajectory of low-pass filter input $v(t)$. Fig. \ref{P5} shows the trajectories of actual system input $u(t)$ and system input with time delay $u(t-\tau)$. And Fig. \ref{P6} shows the adaptation laws of two subsystems.


\section{Conclusion}
According to the results of simulation illustration, the adaptive fuzzy tracking control scheme proposed for the pure-feedback nonlinear system with input delay and full states constraints guarantees that all of the signals in the closed-loop system semi-globally ultimately uniform bounded and the output tracking error can be designed to converge to an arbitrarily small neighbor of origin, while the system full states remain constrained within predefined sets. With the help of Mean Value Theorem, the pure-feedback system is transformed into strict-feedback one. The introduced Pade approximation and the corresponding intermediate are used to eliminate the ill-effects of input delay, while a designed low-pass filter generating actual system input $u(t)$, driven by a newly-defined control input $v(t)$, facilitates the design process of controllers. FLSs are employed to approximate the unknown functions with tunable parameters. With the aid of MLP and DSC technique, the controller design process is simplified. And robust compensators are introduced to circumvent the influences of external disturbance and approximation errors.




\bibliographystyle{IEEEtran}

\end{document}